\documentclass[manuscript]{acmart}
\AtBeginDocument{%
  }
    
\setcopyright{acmcopyright}
\copyrightyear{2026}
\acmYear{2026}

\acmJournal{TOSEM}

\usepackage{enumitem}
\usepackage{subcaption}
\usepackage{placeins}
\usepackage{graphicx}
\graphicspath{{figures/}}
\usepackage{color}
\usepackage{xcolor}
\newcommand{\cmark}{\textcolor{green!60!black}{\checkmark}}
\newcommand{\xmark}{\textcolor{red}{$\times$}}
\usepackage{listings}
\usepackage{dashbox}
\usepackage{booktabs}
\usepackage{soul}
\usepackage{multirow}
\usepackage{multicol}
\usepackage{longtable}
\usepackage{tabularx}
\usepackage{indentfirst}
\usepackage{amsmath}
\usepackage{threeparttable}
\usepackage{tikz}
\usepackage{titlecaps}
\Addlcwords{a, an, and, as, at, but, by, for, in, of, on, or, the, to}
\usepackage{hyperref}

\newcommand{\Ahmed}[1]{}
\newcommand{\yingzhe}[1]{}
\newcommand{\hao}[1]{}
\newcommand{\bram}[1]{}

\newcommand{\hide}[1]{}
\newcommand{\rqone}{What are the structural and maintenance characteristics of the Claude Code plugin marketplaces?}
\newcommand{\rqtwo}{How are plugins developed and maintained?}
\newcommand{\rqthree}{How do plugin components co-evolve?}
\usetikzlibrary{positioning, calc, fit, shapes}

\usepackage[most]{tcolorbox}
\newtcolorbox{summary}[1]{hbox boxed title,
  enhanced,attach boxed title to top left=
    {yshift=-\tcboxedtitleheight/2, yshifttext=-0.5em, xshift=1.5em},
  boxed title style={size=small,colback={black!50!white}},
  title={#1}
}

\begin{document}

\title[An Empirical Study of Claude Code Plugin Marketplaces]{On the Maintenance and Co-evolution of Agent Plugins: An Empirical Study of Claude Code Plugin Marketplaces}

\author{Ahmed~Hereiz}
\email{ahmed.hereiz@queensu.ca}
\affiliation{%
  \institution{Queen's University}
  \department{Software Analysis and Intelligence Lab (SAIL)}
  \city{Kingston}
  \state{ON}
  \country{Canada}
}

\author{Yingzhe~Lyu}
\email{ylyu@cs.queensu.ca}
\affiliation{%
  \institution{Queen's University}
  \department{Software Analysis and Intelligence Lab (SAIL)}
  \city{Kingston}
  \state{ON}
  \country{Canada}
}

\author{Hao~Li}
\email{hao.li@queensu.ca}
\affiliation{%
  \institution{Queen's University}
  \department{Software Analysis and Intelligence Lab (SAIL)}
  \city{Kingston}
  \state{ON}
  \country{Canada}
}

\author{Bram~Adams}
\email{bram.adams@queensu.ca}
\affiliation{%
  \institution{Queen's University}
  \department{Software Analysis and Intelligence Lab (SAIL)}
  \city{Kingston}
  \state{ON}
  \country{Canada}
}

\author{Ahmed~E.~Hassan}
\email{ahmed@cs.queensu.ca}
\affiliation{%
  \institution{Queen's University}
  \department{Software Analysis and Intelligence Lab (SAIL)}
  \city{Kingston}
  \state{ON}
  \country{Canada}
}

\begin{abstract}
AI coding agents, software tools that automate development tasks through reasoning and tool use, are increasingly extended through plugin marketplaces, yet the structure, maintenance, and co-evolution dynamics of these emerging repositories remain empirically unexplored.
Unlike traditional software packages that deliver functionality through source code, agent plugins deliver functionality through a combination of natural-language instruction files, scripts, and configuration files, raising the question of whether these plugins are maintained artifacts that co-evolve across components, or one-off artifacts that developers write once and do not need to revisit.
To study the maintenance and co-evolution of agent plugins, we conduct an empirical study of 1,926 repositories hosting Claude Code plugin marketplaces, analyzing 8,351 plugins and 77,773 commits across 2,018 marketplaces.
We find that the marketplace is expanding rapidly, plugin-touching commit activity growing $8.8\times$ over six months after the October 2025 launch, and plugins targeting Software Engineering tasks accounting for 61.3\% of all plugins.
Plugin development is predominantly feature-driven, with feature commits occurring at more than twice the rate of conventional open-source software (OSS) (39.6\% vs. 17.2\%). Claude co-authors 34.9\% of all commits, and four commit types (\textit{docs}, \textit{perf}, \textit{style}, and \textit{refactor}) carry substantially different meanings in plugin repositories than in traditional software.
Most component types evolve independently, but within skills directories, natural-language instruction files and implementation scripts co-evolve at above-chance rates, with 78\% of co-changes being functionally coupled, representing a new class of maintenance dependency not observed in traditional software engineering.
These findings characterize AI-native plugin repositories as a distinct class of software whose development practices and maintenance dependencies require dedicated empirical frameworks.
\end{abstract}

\ccsdesc[500]{Software and its engineering~Software evolution}
\ccsdesc[500]{Software and its engineering~Maintaining software}
\ccsdesc[300]{Computing methodologies~Artificial intelligence}
\ccsdesc[300]{Software and its engineering~Collaboration in software development}
\keywords{AI Agent, Agentic AI, Agent Plugin, Plugin Marketplace, Claude Code, Skill, MCP}

\maketitle
\section{Introduction}
\label{sec:intro}

Large Language Model (LLM) agents have emerged as autonomous tools for addressing complex tasks that require reasoning, tool usage, and interaction with external environments.
These agents can be extended using plugins, through which developers package and distribute agent behavior, including skills, custom commands, hooks, and model context protocol (MCP) servers, enabling teams to share and reuse consistent agent workflows across projects rather than reconfiguring the agent independently for each project~\citep{claudecode_plugins}.
We use the term \emph{agent plugins}\footnote{\url{https://agent-plugins.org/}} to refer to such distributable behavior packages to distinguish them from traditional software plugins.
Developed by Anthropic, Claude Code was one of the first AI coding agents to support an agent plugin system, though similar capabilities had already been spreading to other AI coding agents.
Agent plugins are distributed through shared repositories called \emph{marketplaces}, which act as catalogs of available plugins that users can register with Claude Code and install from~\citep{claudecode_marketplaces}.

The maintenance and co-evolution of software artifacts are well-studied topics in traditional software engineering~\citep{swanson1976dimensions, zaidman2008coevolution, fluri2007comments}.
Other recent research has extended this focus to adjacent artifact classes, exploring the maintenance health of cross-language bindings for machine learning libraries, identifying technical lags and incomplete release coverage~\citep{li2025bridging}.
Prior work has also investigated the artifact classes of agent skills~\citep{ling2026agentskills}, general-purpose AI assistant stores~\citep{su2025gpt, yan2024chatgpt}, and conventional software package ecosystems~\citep{kikas2017structure, decan2018empirical}.
However, no prior work has studied the maintenance and co-evolution of agent plugins, a new artifact class in which natural-language instruction files and source code artifacts co-exist as primary distributed artifacts.

In multi-component plugins, failing to update co-evolved components together carries the risk of leaving them out of sync, potentially introducing inconsistencies and silent defects.
Unlike traditional software packages, which deliver functionality through source code, Claude Code plugins rely primarily on natural-language Markdown files that Claude reads and interprets at runtime, alongside supporting scripts and configuration files.
This distinction raises the question of whether software engineering methods developed for source-code repositories, which themselves often combine multiple programming languages~\citep{wen2024multilingual}, would transfer to agent plugin repositories.

We hence study the software maintenance and co-evolution of agent plugins, providing the first empirical baseline for this artifact class, and examine whether traditional software engineering methods would transfer to this new artifact class, where natural-language files are the primary distributed artifact rather than source code.
Addressing this issue establishes a foundation for future research into the development patterns and maintenance of this unique artifact class.
Without an empirical baseline for this artifact class, researchers lack the foundation needed to study its development patterns, compare it with conventional ecosystems, or build tools calibrated to its unique characteristics.

We construct our dataset of the Claude Code plugin marketplace by mining agent plugin marketplaces hosted on GitHub. 
Our dataset is the first to capture real-world agent plugin activities across 1,926 repositories, 8,351 plugins, and 77,773 commits.
To study the maintenance and the co-evolution of agent plugins, we investigate the following three research questions~(RQs):
 
\textbf{RQ1: \rqone}
Agent plugin marketplaces have emerged as a key channel for distributing extensions across teams and communities.
However, what types of plugins developers build, whether these marketplaces show signs of active and sustained maintenance, or how their structure compares to traditional package ecosystems is yet to be understood.
Our analysis shows the plugin marketplace is expanding, with plugin-touching commit activity growing $8.8\times$ within six months.
The marketplace is dominated by plugins for software engineering tasks, and 34.4\% of plugins combine multiple component types (e.g., skills paired with hooks or agents).
Our study provides the first empirical baseline for this class of artifact, covering its structural composition, maintenance patterns, and component evolution.

\textbf{RQ2: \rqtwo}
Agent plugin repositories are composed primarily of natural language instructions for LLMs alongside scripts and configuration files, raising the question of whether the same types of development activities that occur in plugin repositories carry the same meaning as in traditional open source software (OSS).
We inspect the development and maintenance activity of commits in agent plugin repositories against the widely accepted Conventional Commits Specification (CCS) definition~\cite{conventionalcommits}.
Our analysis reveals that the actual development and maintenance behavior has shifted from the definition of CCS categories in commits labeled as documentation, performance, style, and refactor under the CCS definition in agent plugin repositories compared to traditional OSS.

\textbf{RQ3: \rqthree}
With 34.4\% of plugins combining multiple component types, whether these components co-evolve during maintenance remains an open question.
Failing to properly update co-evolved components carries the risk of leaving components out of sync, which may introduce inconsistencies and silent defects.
Our results show that most components can be maintained independently.
However, we find that 78\% of the Script--Markdown co-changes within \path{skills/} are functionally coupled, driven by interface changes, internal logic updates, and variable synchronization that propagate from scripts to their paired natural-language instruction files.

This paper makes the following contributions:
\begin{itemize}
    \item We provide the first large-scale empirical characterization of an AI coding agent plugin marketplace, covering structure, development patterns, and component co-evolution across 1,926 repositories, 8,351 plugins, and 77,773 commits.
    \item We show that the developers' intention for commits in agent plugin repositories has shifted from the definition of CCS categories, and provide an approach to reclassify commits by what their diffs actually do, revealing four CCS categories that carry different meanings.
    \item We show that most plugin components evolve independently, but identify a brittle Script--Markdown coupling within skills, and develop a taxonomy of coupling categories.
    \item We release our dataset\footnote{\url{https://github.com/SAILResearch/agentic_plugin_marketplace}} of plugin marketplaces as a foundation for future research on AI-native software development.
\end{itemize}

The remainder of this paper is structured as follows. Section~\ref{sec:background} presents the background and related work. Section~\ref{sec:experimental_design} describes our case study design. Sections~\ref{sec:rq1}--\ref{sec:rq3} address our three research questions. Section~\ref{sec:implications} presents implications for researchers. Section~\ref{sec:threats} discusses threats to validity, and Section~\ref{sec:conclusion} concludes the paper.

\section{Background and Related Work}
\label{sec:background}

In this section, we begin by introducing agent plugin systems, using the Claude Code plugin system as a leading example, followed by a broader discussion of AI coding agents and plugin ecosystems.
Subsequently, we discuss the commit classification for the conventional commits specification (CCS) and the corresponding challenges in the age of AI contribution which we use in our commit analysis in RQ2, as well as software co-evolution studies that are used in our co-change analysis in RQ3. 

\subsection{The Claude Code Plugin System}

Claude Code is an AI-powered, agentic coding tool developed by Anthropic~\citep{claudecode_plugins} that can read codebases, edit files, execute commands, and integrate with development tools.
It supports two ways to add custom skills, agents, and hooks: standalone configuration and plugins~\citep{claudecode_plugins}. 
Standalone configuration relies on the project-level \texttt{.claude/} directory and is typically used for personal workflows. 
In contrast, plugins are designed for reuse across projects, sharing with teammates, and distributing to the community. 
A plugin can contain any combination of supported component types, each serving a distinct purpose. 
Table~\ref{tab:plugin_components} lists all supported types. 

\begin{figure}[t]
\centering
\captionsetup{justification=raggedright,singlelinecheck=false}
\includegraphics[width=\linewidth]{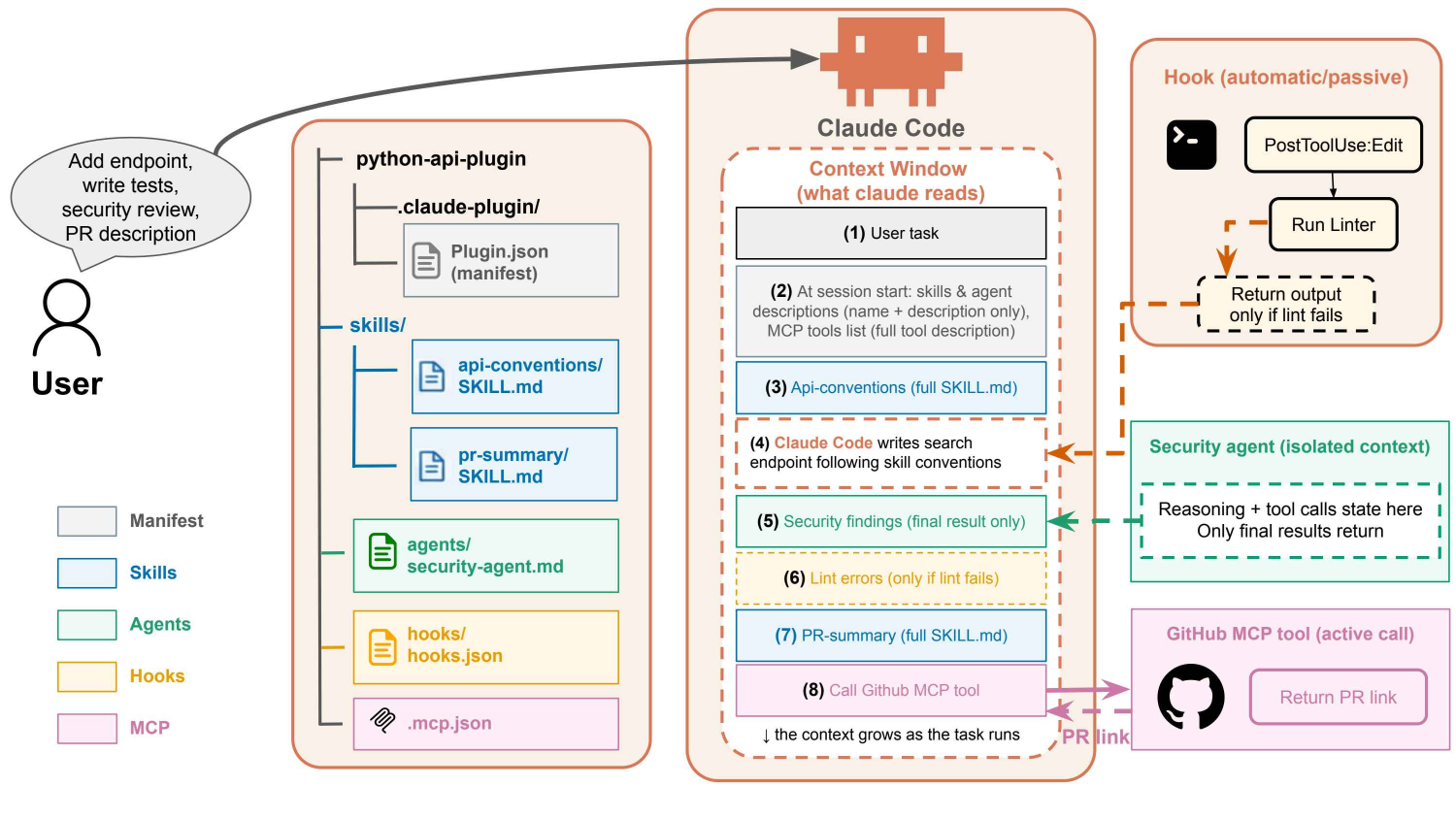}
\caption{Structure and control flow of a Claude Code plugin. The left box illustrates the components in a real plugin (python-api-plugin) installed on disk, and the middle and right boxes illustrate how Claude Code loads and uses the same components over the course of a single task.}
\label{fig:plugin_overview}
\end{figure}

\begin{table}[t]
\centering
\caption{Claude Code plugin component types.}
\label{tab:plugin_components}
\begin{tabular}{lp{10.2cm}}
\toprule
\textbf{Directory / File} & \textbf{Purpose} \\
\midrule
\texttt{commands/}              & User-invoked prompts triggered via slash-commands (e.g., \texttt{/plugin:review})  \\
\texttt{agents/}                & Custom agent definitions \\
\texttt{skills/}                & Agent skills (\texttt{SKILL.md} files) \\
\texttt{hooks/}                 &  Shell commands triggered by events within Claude Code lifecycle (e.g. a \texttt{PostToolUse} hook was triggered to run a linter once Claude edits a file)\\
\texttt{.mcp.json}              & Model Context Protocol (MCP) server configurations \\
\texttt{.lsp.json}              & Language Server Protocol (LSP) server configurations \\
\texttt{settings.json}          & Default Claude Code configuration \\
\texttt{.claude-plugin/plugin.json} & Optional plugin manifest \\
\bottomrule
\end{tabular}
\end{table}

Three component types shape Claude's responses during a session.
\emph{Commands} (\texttt{commands/}) are Markdown files users call as slash-commands
(e.g., \texttt{/plugin-name:review}), each defining a reusable prompt for a single bounded interaction.
\emph{Skills} (\texttt{SKILL.md} files in named subdirectories of \texttt{skills/}) are primarily model-invoked, which means that Claude selects and applies them automatically based on the current task, though users can also invoke a skill explicitly.
\emph{Agents} (\texttt{agents/}) are also model-invoked, but each runs in its own context
window with a custom system prompt and restricted tool access, so Claude can carry out a
subtask in isolation rather than accumulating it in the main conversation.

The four remaining types operate independently of the conversation, triggered by session events or applied as configuration rather than selected by Claude during a task.
\emph{Hooks} (\texttt{hooks/hooks.json}) are shell commands that run automatically at fixed points in Claude Code's lifecycle~\citep{claudecode_hooks}, giving deterministic control over its behavior instead of relying on the model to decide whether to act, for example, running a linter after every file write.
\emph{MCP servers} (\texttt{.mcp.json}) connect Claude to external tools and data sources through the MCP~\citep{mcp2024}, giving it direct access to databases, issue trackers, and APIs without requiring the user to copy data into chat.
\emph{LSP servers} (\texttt{.lsp.json}) connect Claude to LSP daemons~\citep{lsp2016} for real-time code intelligence such as go-to-definition and type checking.
\texttt{settings.json} supplies default configuration applied whenever the plugin is enabled. Currently, it supports only two settings: which of the plugin's agents to promote to the main conversation thread, and the status line to show for that agent.

Figure~\ref{fig:plugin_overview} illustrates the loading and execution process for a single task. The user asks Claude Code to add an endpoint, write tests, run a security review, and open a pull request, placing that task in Claude's context window. Because \texttt{api-conventions} matches the task, its full \texttt{SKILL.md} loads in step~3 (only its name and description were loaded in step~2), and Claude writes the endpoint following its conventions. The edit fires a hook that runs a linter, whose output enters the context only if it fails. Claude delegates the security review to \texttt{security-agent}, which reasons and calls tools inside its own isolated context window and returns only its final result. The \texttt{pr-summary} skill loads the same way \texttt{api-conventions} did, and Claude calls \texttt{github-mcp-server} to open the pull request, whose link is added to the context. Claude reports the security findings and the pull request link back to the user.

Agent plugin distribution is facilitated through \emph{marketplaces}~\citep{claudecode_marketplaces}, which act as catalogs of available plugins defined via a \texttt{marketplace.json} file. Registering a marketplace with Claude Code, then installing individual plugins from it are the two steps required to use third-party plugins~\citep{claudecode_discover}.
An official marketplace is automatically registered by default, users can additionally register third-party marketplaces hosted on GitHub or other Git providers, or local directories.

\subsection{Software Package Ecosystems}

Empirical software engineering has a long tradition of studying software package ecosystems,  which we extend to the emerging class of agent plugin marketplaces.

Several studies have analyzed the structure and evolution of conventional software package registries through the lens of dependency networks.
\citet{kikas2017structure} analyzed dependency networks across npm, RubyGems, and Cargo, finding that transitive dependencies grow rapidly and that significant structural differences exist across ecosystems.
\citet{decan2018empirical} extended this comparison to seven packaging ecosystems, collectively hosting more than 830,000 packages and 5.8 million releases, finding that all ecosystems grow continuously over time in both package count and update frequency.
While these studies focus on dependency relationships between packages, we study the Claude Code agent plugin marketplaces as a new class of AI-native software, where distributed artifacts are natural-language instruction files, scripts, and configuration files rather than compiled packages.

\citet{onagh2025extension} studied extension decisions in the GitHub Marketplace, analyzing 6,983 CI Actions from 3,869 providers and finding that the marketplace expands by approximately 41\% annually and that 65\% of newly released CI Actions replicate existing capabilities within six months of the original tool's launch.
While Onagh and Nayebi studied functional replication in a non-AI code-based extension marketplace (GitHub CI Actions), we study the agent plugin marketplaces, where the artifacts are natural-language instruction files rather than executable actions.

\citet{hassan2024fmware} introduced the concept of \emph{FMware}, software that relies on foundation models as a core component, and catalogued the engineering challenges that arise when natural language becomes a primary software artifact.
\citet{li2025promptmgmt} built on this by empirically studying 24,800 open-source prompts from 92 GitHub repositories, introducing \emph{promptware} as a specific instance of FMware built using natural language prompts.
The authors find that prompts are predominantly stored in Markdown (72.8\%), with 38.5\% exhibiting semantic duplication and 80.1\% falling below standard reading-ease thresholds, showing that natural-language software assets face maintenance challenges analogous to those in traditional code.
Claude Code agent plugins are one instance of promptware, distinguished by co-existing with executable scripts and configuration files that co-evolve within the same repository.

Several studies have empirically analyzed general-purpose AI assistant plugin and extension ecosystems.
\citet{su2025gpt} mined 722,349 GPTs from the GPT Store, analyzing the platform's classification system, interaction modes, and user behavior at scale. GPTs are built through natural language instructions and knowledge files.
\citet{yan2024chatgpt} conducted the first comprehensive characterization of the ChatGPT plugin store, classifying 1,038 plugins into 21 functional categories using zero-shot NLI classification. ChatGPT plugins are API integrations described by a JSON manifest file.
They found an uneven distribution of functionality, with more than half of all plugins concentrated in five categories: Data \& Research (12.9\%), Tools (11.2\%), Business (10.1\%), Developer \& Code (9.7\%), and Entertainment (6.7\%).
While Su et al.\ and Yan et al.\ studied AI assistant stores whose artifacts carry no versioned development history, Claude Code agent plugins are GitHub repositories combining natural-language instruction files, executable scripts, and configuration files, enabling commit-level analysis of development practices and component co-evolution.

\citet{ling2026agentskills} conducted a large-scale analysis of 40,285 agent skills from the \texttt{skills.sh} marketplace, finding that skills concentrate heavily in software engineering workflows with widespread intent-level redundancy.
\citet{zhu2026skillclone} extended this line of work by detecting multi-modal clone relationships across 20,000 skills, finding that 75\% of skills participate in at least one clone pair and that the ecosystem is inflated 3.5$\times$ by undeclared reuse, with 40\% of clone pairs crossing author boundaries.
While Ling et al.\ and Zhu et al.\ studied skills artifacts, we study the Claude Code plugin marketplace, where skills are one component type within a structured multi-component plugin system distributed through a dedicated marketplace.  Table~\ref{tab:study_comparison} summarizes the differences between these studies and ours.

\subsection{Commit Classification and AI Contributions}

Before the Conventional Commits Specification, researchers proposed manual taxonomies for classifying software changes. \citet{swanson1976dimensions} introduced one of the earliest taxonomies, distinguishing corrective, adaptive, and perfective maintenance. \citet{hindle2008large} extended this by manually classifying 2,000 large commits across nine OSS projects, introducing additional categories including Implementation, Module Management, and Non-functional changes, and finding that large commits are more perfective while small commits are more corrective. \citet{bhatia2023towards} further extended Hindle et al.'s taxonomy to ML research repositories, finding that ML-specific artifacts require two new top-level categories (\textit{Data} and \textit{Dependency Management}) and 16 new sub-categories, showing that commit taxonomies need adaptation when the primary artifact shifts away from traditional source code.

The conventional commits specification (CCS)~\citep{conventionalcommits} standardized 11 commit types, providing a machine-readable schema for communicating the intent of a change that is increasingly adopted in open-source projects.
\citet{zeng2025ccs} studied 88,704 commits from 116 traditional open-source projects that explicitly adopt CCS, finding that \textit{fix} and \textit{chore} together account for 53\% of commits and that developers frequently misuse CCS labels due to overlapping type definitions.
While \citet{zeng2025ccs} studied CCS adoption in traditional open-source projects, we study commit type distributions in the plugin marketplace, where the primary artifact is a natural-language instruction file rather than source code.

\citet{honel2020density} improved automated commit classification by combining commit message keywords with source code density metrics.
\citet{wan2025commitsuite} established a comprehensive benchmark for CCS commit classification and message generation, providing a standardized evaluation framework for automated classification methods.
\citet{li2025rise} introduced AIDev, a large-scale dataset of 456,535 agent-authored pull requests from 61,453 repositories, and applied LLM-based CCS annotation at scale to label each PR's purpose, demonstrating that automated CCS classification is viable for large collections of unlabeled commits.
\citet{li2026aidev} published a dedicated dataset paper with 932,791 Agentic pull requests from 116,211 repositories, providing formal infrastructure for large-scale studies of AI coding agent behavior.
While prior work developed and validated commit classification methods for traditional source-code repositories, we apply commit classification to the agent plugin marketplaces, where the primary artifact is a natural-language instruction file rather than compiled code, and during manual validation we inspect file-level diffs alongside commit messages.

Researchers have studied AI contributions in traditional source-code repositories. \citet{tufano2024chatgpt} mined ChatGPT mentions across open-source projects, identifying 467 confirmed instances of AI-assisted contributions and classifying them into 45 tasks spanning documentation, refactoring, test generation, and feature implementation.
\citet{robbes2026promises} generalized detection beyond explicit mentions, cataloguing co-author trailers, email patterns, and branch prefixes across a range of coding agents as heuristics for identifying agent contributions in commit histories.
In this paper, we apply co-authorship detection heuristics to the plugin marketplace.

In addition, \citet{watanabe2025agentic} studied 567 Claude Code pull requests on GitHub, finding that 83.8\% were accepted by project maintainers, establishing that agentic contributions integrate at rates comparable to human contributions.
\citet{horikawa2025refactoring} showed that agentic refactoring is dominated by low-level, consistency-oriented edits rather than the high-level design changes common in human refactoring.
\citet{ouatiti2026logging} found that AI coding agents log less frequently than humans (58.4\% of repositories), but at higher density when they do, with 67\% of generated logging statements non-compliant with project logging instructions and 72.5\% of those requiring repair by human developers.
These studies operate exclusively on source-code repositories. In our study, we examine how these behavioral patterns manifest in an AI-native marketplace where the primary artifact is a natural-language instruction file rather than compiled code.

\subsection{Software Co-evolution}

A body of empirical work studies co-evolution, the tendency of different artifact types within a project to change together. Such co-evolution typically indicates extra maintenance work, and forgetting to update either co-evolving artifact increases the risk of breakage.
\citet{mcintosh2011build} first applied association rules to measure co-evolution between build files and source code in ten C and Java projects, finding that build maintenance imposes up to 27\% overhead on production code development.
\citet{jiang2015coevolution} applied association rules to measure co-evolution between infrastructure-as-code files and source code across OpenStack projects, finding tight coupling between infrastructure and both production and test files. \citet{barrak2021coevolution} extended this to machine-learning projects, finding a median co-change confidence of 91.91\% between DVC pipeline files and source code.
While McIntosh et al., Jiang and Adams and Barrak et al.\ applied association-rule frameworks to well-defined artifact pairs in source-code repositories (infrastructure--source and pipeline--source), we apply the same framework to AI-native plugin repositories, where the co-evolution relationships are between natural-language component types rather than between code and its companion artifacts.

\citet{zaidman2008coevolution} mined the version histories of two open-source systems alongside test coverage reports, identifying three growth patterns in how production and test code co-evolve: synchronous, time-delay, and test backlog. \citet{marsavina2014coevolution} studied five open-source systems using Apriori association rule mining (support 50\%, confidence 60\%), identifying six fine-grained co-evolution patterns at the class and source-code entity level. \citet{fluri2007comments} examined three OSS systems (ArgoUML, Azureus, and JDT Core), finding that 97\% of comment updates occur in the same revision as the associated code change. These studies examine co-evolution between code and its companion artifacts (tests or comments). 
While prior studies examined co-evolution between code and its companion artifacts (tests or comments) in traditional OSS, we examine co-evolution within AI-native plugin repositories between implementation scripts and their paired natural-language instruction files.

\begin{table}[t]
\centering
\caption{Comparison of related AI ecosystem studies. NL = natural language.}
\label{tab:study_comparison}
\begin{tabular}{lcccc}
\toprule
& \textbf{Su et al.}~\citep{su2025gpt} & \textbf{Yan et al.}~\citep{yan2024chatgpt} & \textbf{Ling et al.}~\citep{ling2026agentskills} & \textbf{Our study} \\
\midrule
\textbf{Target system}         & Conversational LLM  & Conversational LLM  & AI coding agent      & AI coding agent          \\
\textbf{Scale}                 & 722,349 GPTs        & 1,038 plugins       & 40,285 skills        & 8,351 plugins            \\
\textbf{Artifact form}         & NL customization    & API integration & NL skill file        & Multi-component package  \\
\textbf{NL instruction files}  & \cmark          & \xmark            & \cmark           & \cmark               \\
\textbf{Executable scripts}    & \xmark          & \xmark            & \xmark           & \cmark               \\
\textbf{Config files}          & \xmark          & \cmark            & \xmark           & \cmark               \\
\textbf{Repository-based}      & \xmark          & \xmark            & \cmark           & \cmark               \\
\bottomrule
\end{tabular}
\end{table}

\section{Data Collection}
\label{sec:experimental_design}
\begin{figure}[t]
\centering
\includegraphics[width=\textwidth]{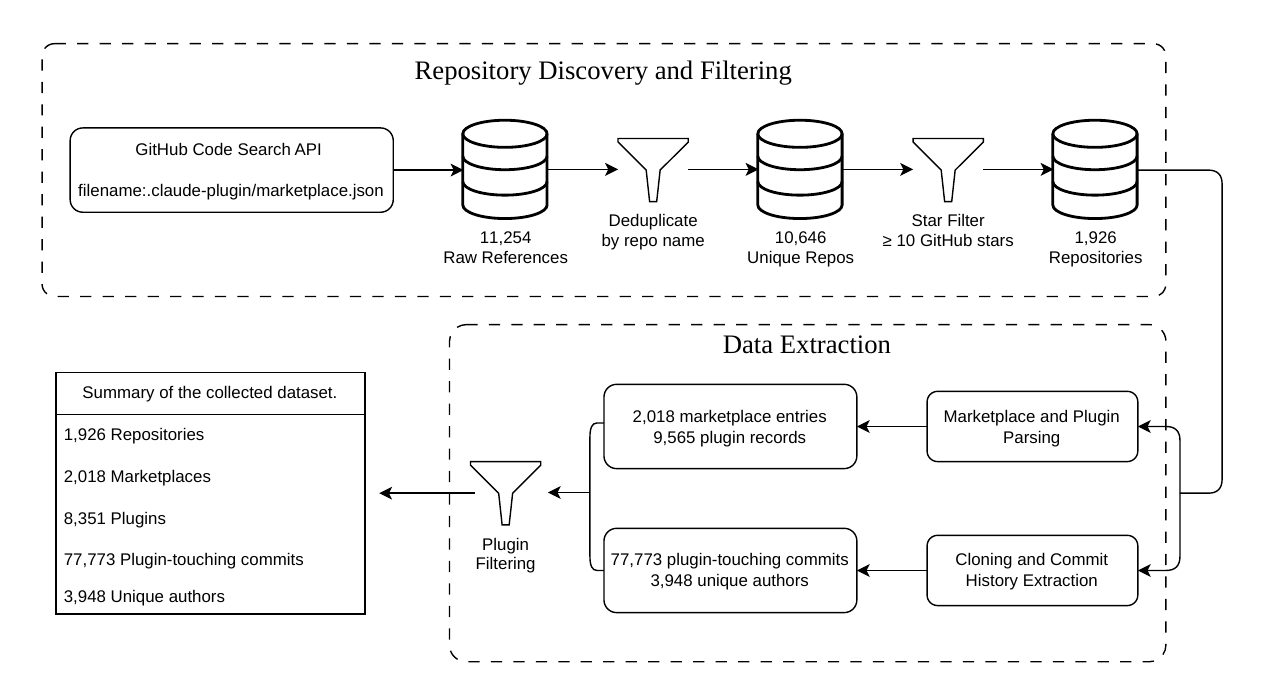}
\caption{Overview of the data collection pipeline and dataset summary.}
\label{fig:pipeline}
\end{figure}

\subsection{Repository Discovery and Filtering}
\label{sec:discovery}

Claude Code uses a publicly documented schema~(\texttt{marketplace.json})~\citep{claudecode_plugins} that is hosted on GitHub, making the full ecosystem discoverable via the GitHub Code Search API and enabling the large-scale empirical analysis we conduct in our RQs.
To support the empirical analyses for studying the marketplace characteristics, development and maintenance activities, and the component co-evolution. 
Our data collection follows a two-step pipeline: 1) repository discovery and filtering, and 2) data extraction.
Figure~\ref{fig:pipeline} illustrates the pipeline and also provides a summary of the resulting dataset.
To identify all GitHub repositories that host a Claude Code plugin marketplace, we use the GitHub Code Search API with the query \path{filename:.claude-plugin/marketplace.json} to discover all repositories containing the manifest file required for a Claude Code plugin marketplace.
We apply a divide-and-conquer strategy to circumvent the 1000 results per query limit on GitHub's Code Search API, as done in the prior work~\citep{kalliamvakou2016github}.
The search was executed on 2 April 2026 and yielded 11,254 raw repository references.
The raw search results contain duplicate references arising from identical files indexed under different paths or API pages.
We deduplicate by repository full name, retaining one record per unique repository, reducing the 11,254 raw references to 10,646 unique repositories.

Many repositories in the raw set are personal experiments, test repositories, or forks with no community traction.
To retain only repositories that represent community-adopted plugins, we apply a minimum threshold of 10 GitHub stars, following prior MSR studies~\citep{nagappan2013msr}. Plugins are a newer artifact class than the software projects these thresholds were calibrated on, so we also evaluated the analysis at 5 and 25 stars.
As we discuss in the Threats to Validity (Section~\ref{sec:threats}), these qualitative findings are stable across all three thresholds.
Out of the 10,646 unique repositories, 81.7\% have fewer than 10 stars, including 46.3\% that carry zero stars, confirming that a threshold is necessary to separate low-quality noise from community-adopted repositories.

\subsection{Data Extraction}
\label{sec:extraction}

\textbf{Marketplace and Plugin Parsing.}
To obtain plugin-level records (plugins declared inside \texttt{marketplace.json} files), we parse each marketplace file found in the cloned repositories following the Anthropic plugin marketplace schema~\citep{claudecode_plugins}. 
Parsing the cloned repositories yields 2,018 marketplaces and 9,565 raw plugins.
We then apply a filtering step to remove entries whose \texttt{source} field references a plugin located outside the marketplace repository rather than a local subdirectory, as we can only analyze plugins whose files are directly available in the cloned repository.
We also filter out entries that are cross-marketplace duplicates, true duplicates, entries with a blank source field, and entries whose declared path does not resolve to an existing directory in the cloned repository for the same reason.
After filtering, 8,351 valid, locally-resolvable plugins remain.

\textbf{Cloning and Commit History Extraction.}
We clone all 1,926 repositories and extract their complete Git commit histories from the cloned repositories.
We record the commit-level metadata including author, timestamp, commit title, commit message, and the file changes for each commit that touches plugin component files shown in Table~\ref{tab:plugin_components}.
We focus only on plugin-touching commits rather than all commits in the repository because we want to study how plugin components are developed and maintained.
Commits that do not touch any plugin component reflect broader repository activity outside our scope. This process yields 77,773 plugin-touching commits (i.e., commits that modify at least one plugin component file, regardless of whether other non-plugin files are also changed) contributed by 3,948 unique authors across 1,926 repositories.
The median number of plugin-touching commits per repository is 13 (mean = 40.4), and only 115 repositories (6.0\%) contain a single commit, indicating that the majority of repositories represent ongoing development rather than one-off uploads.

\section{RQ1: \titlecap{\rqone}}
\label{sec:rq1}
\subsection{Motivation}

Prior work has shown that natural language is becoming a primary software artifact with distinct engineering challenges~\citep{hassan2024fmware, li2025promptmgmt}.
Claude Code plugins represent one of the first large-scale software ecosystems built primarily around natural-language artifacts.
These artifacts range from Markdown instruction files that AI agents read at runtime, executable scripts and configuration files, yet their growth trajectory, functionality of the plugins, structural composition, and development activity remain unstudied.
Understanding whether the development activities of agent plugin marketplaces are accelerating or represent only an initial burst reveals whether the marketplace has moved beyond an initial adoption phase into sustained development. Prior work on software ecosystems has shown that a significant share of projects die in their first year, with fewer than half surviving five years~\citep{ait2022survival}.
Sustained growth motivates understanding what developers are actually building, specifically the functionality of these plugins, and the kinds of tasks being delegated to AI agents.
Knowing what plugins do, their internal structure in terms of which component types developers use and how those types co-occur identifies the technical artifacts developers must account for when building and maintaining plugins.
Plugin maintenance and update patterns reveal whether contributors are actively improving their work over time or abandoning plugins after an initial release.
Taken together, these four dimensions offer researchers one of the first large-scale empirical windows into how developers build and maintain software whose primary deliverable is natural-language instructions, executable scripts, and configuration files.

\subsection{Approach}

To study the agent plugin marketplace, we use the dataset described in Section~\ref{sec:extraction} and conduct four analyses. 

\textbf{Trace repository and component creation.}
We retrieve each repository's creation date from the GitHub REST API across all 1,926 repositories and use it as the date the marketplace entered the ecosystem.
To obtain a finer-grained, component-level view, we identify the earliest commit that introduced each component file: \path{SKILL.md} files within \path{skills/} subdirectories, \path{.md} command files within \path{commands/}, \path{.md} agent definitions within \path{agents/}, hook configuration files within \path{hooks/}, \path{plugin.json} manifests, \path{marketplace.json} registries, and \path{.mcp.json} server configurations.
We then perform a non-parametric Mann-Whitney U test~\cite{mann1947test, wilcoxon1945individual}, which makes no assumption of normal distribution, to check if there is a significant difference in the monthly repository-creation counts between the pre-launch and post-launch periods.
We also apply Cliff's $\delta$~\cite{cliff1993dominance} effect size to measure the magnitude of difference between two groups of observations.
We apply the thresholds provided by \citet{romano2006appropriate} for Cliff's $\delta$: 

\begin{equation} \label{effectsize}
\mathrm{Effect \ size} = 
\left\{
\begin{array}{ll}
	negligible,  & \mathrm{if} \ |\delta|  \le 0.147 \\
	small,  & \mathrm{if} \ 0.147 < |\delta|  \le 0.33 \\
	medium,  & \mathrm{if} \ 0.33 < |\delta|  \le 0.474 \\
	large,  & \mathrm{if} \ 0.474 < |\delta|  \le 1 \\
\end{array}\right.
\end{equation}

\textbf{Classify plugin functionality.}
To understand what developers are actually building and the functionality of these plugins, we classify all 8,351 valid plugins (Section~\ref{sec:extraction}) into a two-layer taxonomy of 6 major categories and 20 sub-categories, adopting the same classification scheme used by \citet{ling2026agentskills} for agent skills.
The categories proposed by \citet{ling2026agentskills} cover end-to-end agent workflows in categories like Software Engineering, Information Retrieval, and Content Creation. We verified after applying the taxonomy that no new categories were required, confirming that it transfers to Claude Code plugins without modification.
Given a plugin's name and description, we label plugins with Qwen3-Coder-Next-80B from the same Qwen family used by \citet{ling2026agentskills} (Qwen2.5-32B-Instruct), assigning a plugin to one sub-category with the plugin name and description as input.
The prompt template we used for classifying plugins is provided in Appendix~\ref{app:classification_prompt}.

\textbf{Analyze plugin composition.}
We apply a four-level approach across the 2,018 marketplaces and 8,351 valid plugins described in Section~\ref{sec:extraction}.
At the marketplace level, we count the number of plugins declared in each \path{marketplace.json}.
At the plugin level, we check whether each plugin root contains any of the seven component types defined in the Claude Code documentation~\citep{claudecode_plugins}: \path{commands/}, \path{agents/}, \path{skills/}, \path{hooks/}, \path{.mcp.json}, \path{.lsp.json}, and \path{settings.json}.
We then construct a co-occurrence matrix recording which component types appear together within the same repository, and we group files by extension to characterize the file types each component uses within each component directory.

\textbf{Measure commit activity growth.}
To understand the plugin development and update patterns, we count plugin-touching commits per month across the 72,191 plugin-touching commits from the official launch of the Claude Code plugin marketplace in October 2025 through the data cutoff at the end of March 2026, using the commit timestamps from the dataset described in Section~\ref{sec:extraction}.
We perform the non-parametric Mann-Kendall test~\cite{mann1945nonparametric, kendall1962rank} to determine if there is a monotonic upward or downward trend in monthly commit count.
The null hypothesis (H0) for the Mann-Kendall test is that there is no monotonic trend across our ordered sequence of six monthly commit counts and the alternative hypothesis (H1) is that a monotonic trend is present. 
We also apply the Sen's slope~\cite{sen1968estimates} to estimate the magnitude if a statistically significant Mann-Kendall correlation is present.
Sen's slope is a non-parametric estimator that indicates the rate of change per unit time step.
Following the same aggregation-based mitigation for serial autocorrelation (where consecutive monthly observations tend to be correlated with each other) used by~\citet{kudrjavets2023codevelocity}, we compute both tests on monthly-aggregated commit counts rather than raw daily timestamps.

\subsection{Results}

\begin{table}[t]
    \centering
    \caption{Taxonomy and category-level statistics of agent plugins.}
    \label{tab:rq1_taxonomy}
    \begin{tabular}{llrrrr}
    \toprule
    \textbf{Major Category} & \textbf{Sub-category} & \textbf{\# Plugins} & \textbf{\% of Total} & \textbf{\% skills.sh}~\citep{ling2026agentskills} & \textbf{\# Median Stars}\\
    \midrule
    Software Engineering
    & Code Generation & 2,164 & 25.9\% & 14.3\% & 40.0 \\
    & Infrastructure & 1,387 & 16.6\% & 24.0\% & 40.0 \\
    & Debug \& Analysis & 1,234 & 14.8\% & 13.2\% & 49.0 \\
    & Version Control & 336 & 4.0\% & 3.2\% & 37.5 \\
    \midrule
    Information Retrieval
    & Academic Search & 196 & 2.3\% & 2.7\% & 46.5 \\
    & Web Search & 133 & 1.6\% & 1.4\% & 47.5 \\
    & Live Data Streams & 24 & 0.3\% & 0.7\% & 68.0 \\
    \midrule
    Productivity Tools
    & Task Management & 481 & 5.8\% & 5.6\% & 33.0 \\
    & Document Systems & 275 & 3.3\% & 3.9\% & 46.5 \\
    & Team Communication & 207 & 2.5\% & 1.7\% & 40.0 \\
    \midrule
    Data \& Analytics
    & Data Processing & 468 & 5.6\% & 7.9\% & 40.5 \\
    & Data Visualization & 130 & 1.6\% & 1.8\% & 30.5 \\
    & Math \& Calculation & 70 & 0.8\% & 0.9\% & 43.5 \\
    \midrule
    Content Creation
    & Text Generation & 430 & 5.1\% & 5.5\% & 44.0 \\
    & Image Generation & 146 & 1.7\% & 3.0\% & 52.0 \\
    & Audio \& Video & 133 & 1.6\% & 3.6\% & 47.5 \\
    \midrule
    Utilities \& Other
    & Other Utilities & 218 & 2.6\% & 2.8\% & 41.0 \\
    & Memory \& Cognition & 208 & 2.5\% & 2.3\% & 48.5 \\
    & Command Execution & 62 & 0.7\% & 0.8\% & 47.5 \\
    & Local File Control & 24 & 0.3\% & 0.6\% & 113.0 \\
    \bottomrule
    \end{tabular}
\end{table}

\textbf{Plugins designed for software engineering tasks dominate the plugin marketplaces, with their four sub-categories collectively accounting for 61.3\% of all plugins, far ahead of any other major category.}
Table~\ref{tab:rq1_taxonomy} shows the full distribution across six major categories and 20 sub-categories, alongside the corresponding percentages from \citet{ling2026agentskills} for comparison.
This concentration mirrors the distribution reported by \citet{ling2026agentskills} for the \texttt{skills.sh} agent skills marketplace, suggesting that the Software Engineering dominance is not specific to this marketplace but reflects the broader early-adopter profile of AI coding agent platforms, where the primary user base consists of software developers extending their own workflows.
The remaining five categories all fall within 2 percentage points of the \texttt{skills.sh} distribution, except Content Creation (8.2\% vs.\ 12.1\%) and Data \& Analytics (7.9\% vs.\ 10.6\%), which are moderately lower in our study.

\begin{figure}[t]
  \centering
  \includegraphics[width=\linewidth]{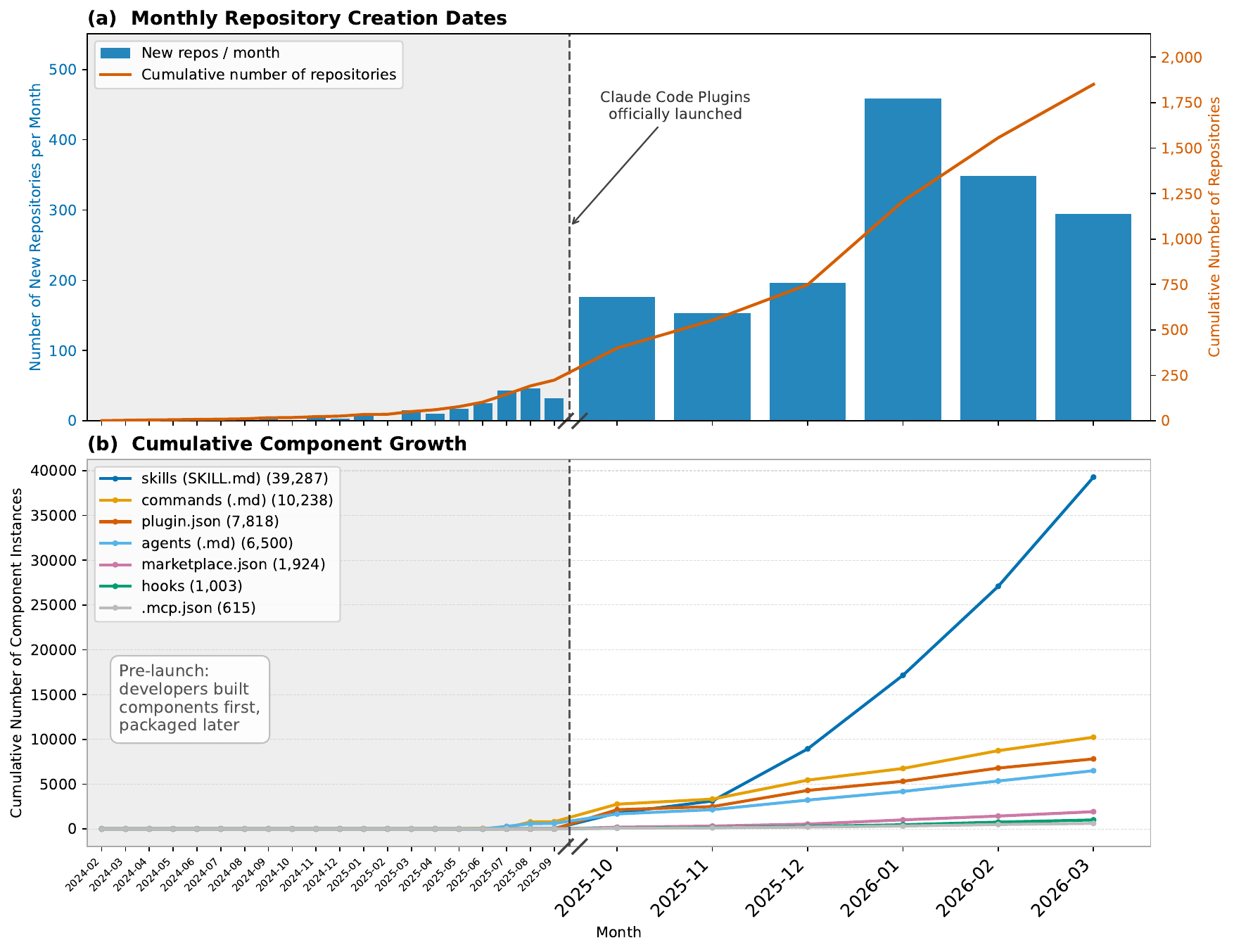}
  \caption{Plugin marketplace repository age and component growth.
           The x-axis uses compressed spacing for the pre-launch period
           (January~2024--September~2025) and uniform spacing thereafter,
           axis-break marks signal the scale change.
           \textbf{(a)}~Monthly distribution of plugin marketplace repository creation dates, with 84.5\% created in the six months following the October~2025 launch.
           \textbf{(b)}~Cumulative plugin component counts by type. All component types grow over time.}
  \label{fig:rq0_growth_combined}
\end{figure}

\textbf{Repository creation in the second half of the post-launch period (Jan--Mar 2026) was $2.1\times$ higher than in the first half (Oct--Dec 2025), indicating accelerating adoption beyond the initial launch.}
Figure~\ref{fig:rq0_growth_combined}(a) shows that 1,102 new repositories were created during Jan--Mar 2026, compared with 525 during Oct--Dec 2025, representing a $2.1\times$ increase.
The majority of these repositories (84.5\%) were newly created during the six-month post-launch window, while the remaining 15.5\% were pre-existing repositories whose owners registered a plugin manifest after the standards were announced.
The Mann-Whitney U test shows that the difference in monthly repository creation counts between the 20-month pre-launch period and the 6-month post-launch period is statistically significant~($p < 0.001$), with Cliff's $\delta = 1.0$ indicating a large magnitude of the statistical difference.
These statistical test results confirm that repository creation concentrates in the post-launch period.

\textbf{Skill instances grew at a substantially higher rate than all other component types, with the gap widening further from commands and agents from November 2025 onwards.}
Figure~\ref{fig:rq0_growth_combined}(b) shows that skills diverged from the other component types starting November 2025, growing 22$\times$ from 1,776 instances at launch to 39,287 by March 2026, compared to commands growing 3.7$\times$ (2,763 to 10,238) and agents 3.9$\times$ (1,671 to 6,500).
By March 2026, skills instances (39,287) outnumbered all other component types combined (20,280), with skills consisting primarily of natural-language instruction files, reflecting natural language as the dominant artifact type across the marketplace.

\begin{figure}[t]
    \centering
    \includegraphics[width=\linewidth]{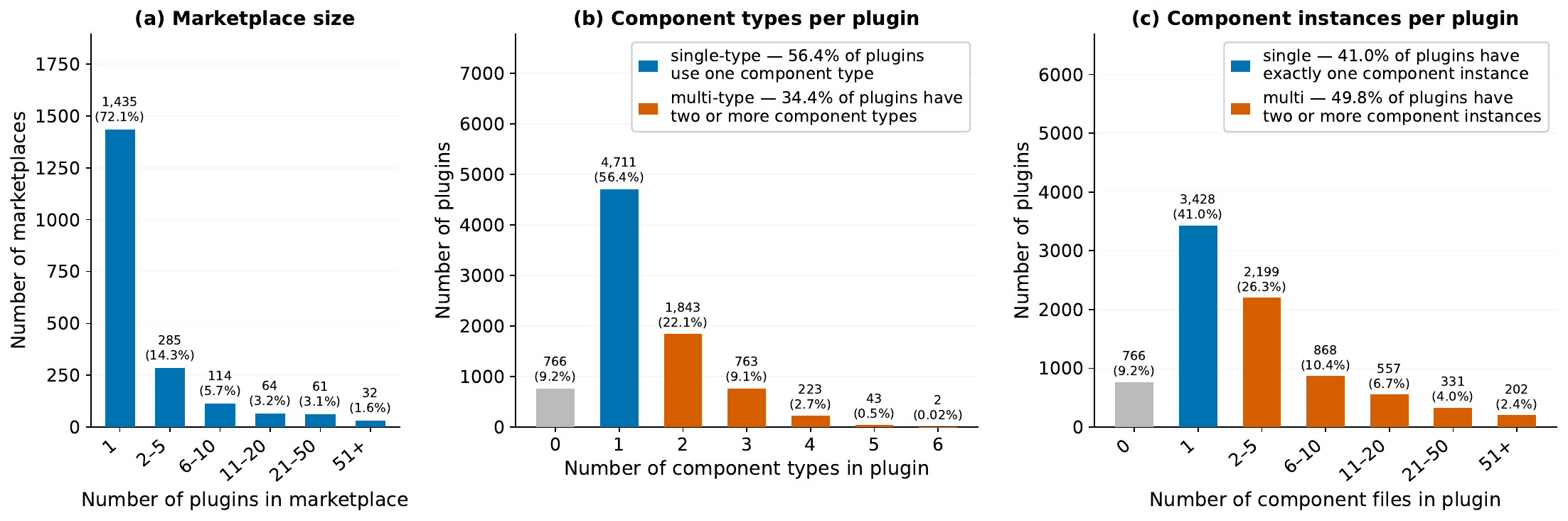}
    \caption{(a) Distribution of marketplace sizes (number of plugins per \texttt{marketplace.json}).
      (b) number of distinct component types per plugin.
      (c) total number of component files per plugin.}
    \label{fig:rq0_marketplace_sizes}
\end{figure}

\textbf{34.4\% of plugins combine two or more component types, and 72.1\% of marketplaces host a single plugin.}
Figure~\ref{fig:rq0_marketplace_sizes} shows that all three distributions are heavily right-skewed, with 56.4\% (4,711 of 8,351) of plugins relying on a single component type while 49.8\% (4,157 of 8,351) contain two or more component file instances. \path{sangrokjung/claude-forge} exemplifies the upper end of structural diversity, combining 6 of the 7 component types within a single plugin, including 56 commands and 11 agents alongside skills, hooks, rules, and a settings manifest. 
The single-plugin marketplace majority reflects individual repositories publishing self-contained extensions for personal or team use, while at the tail, three community aggregators alone account for 12.1\% of all plugins: \path{jeremylongshore/claude-code-plugins-plus-skills} (417), \path{TheBushidoCollective/han} (338), and \path{BbgnsurfTech/claude-skills-collection} (258), each indexing contributions from multiple sources into a single registry.

\begin{figure}[t]
  \centering
  \includegraphics[width=\linewidth]{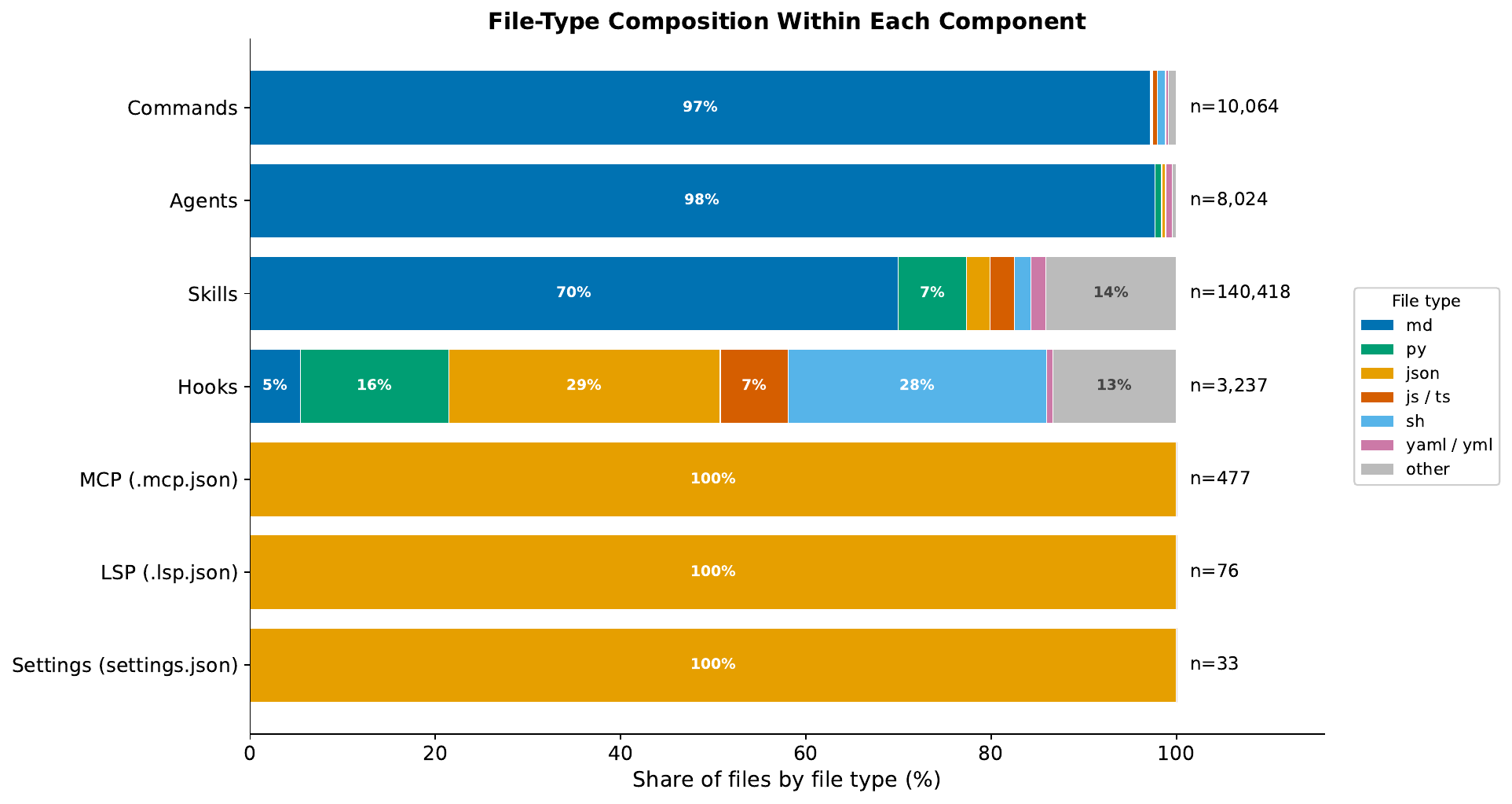}
  \caption{File type composition within each plugin component type.
           Instruction-oriented components (Commands, Agents, Skills)
           are predominantly Markdown, while configuration-oriented
           components (MCP, LSP, Settings) are exclusively JSON.}
  \label{fig:rq0_file_type_composition}
\end{figure}

\textbf{Instruction-oriented components are written almost entirely in Markdown, while configuration-oriented components are exclusively JSON.}
Figure~\ref{fig:rq0_file_type_composition} shows how file type varies with the role each component plays in the plugin architecture.
Commands, agents, and skills are predominantly Markdown, specifying plugin behavior by writing natural language instructions in \texttt{.md} files.
Model context protocol (MCP), Language server protocol (LSP) and settings components are exclusively JSON, reflecting their role as structured tool and environment configuration. 
Skills and hooks are the two mixed cases. Skills occasionally include non-Markdown files alongside their primary instruction content, while hooks are the most heterogeneous component overall, combining shell scripts, Python, and JavaScript files with JSON configuration.
The small non-Markdown share observed for \texttt{commands/} (\textless 3\%) reflects leftover files left in the directory rather than a second component format. Claude Code only reads the \texttt{.md} file when invoking a command, consistent with commands being defined as single Markdown files in the Claude Code documentation~\citep{claudecode_plugins}.

\begin{figure}[t]
  \centering
  \includegraphics[width=0.6\linewidth]{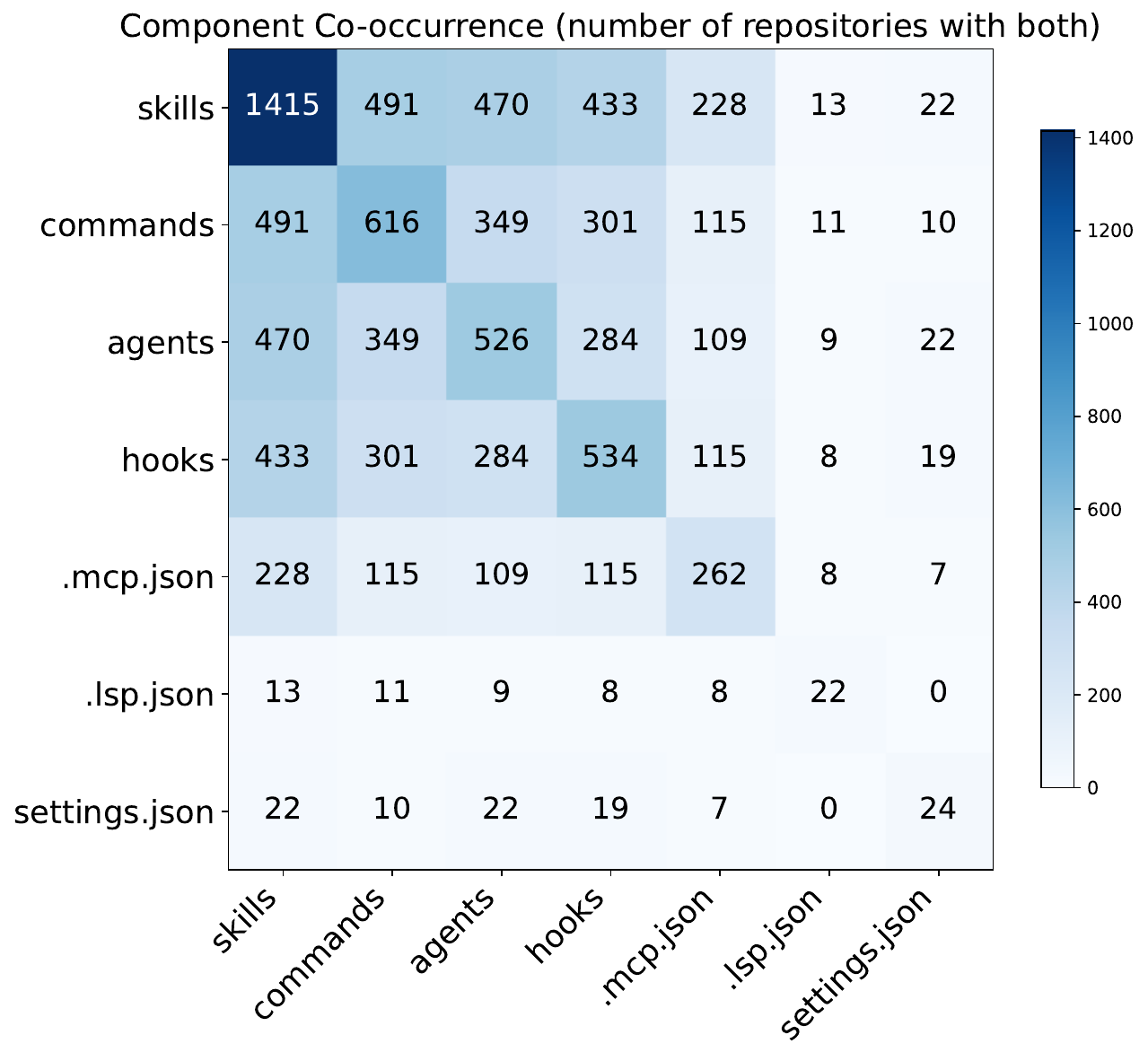}
  \caption{Component co-occurrence matrix across repositories.
           Each cell shows the number of repositories that contain
           both components. Skills co-occur most frequently with
           all other component types due to their dominant presence.}
  \label{fig:rq0_component_cooccurrence}
\end{figure}

\textbf{Among plugin repos that combine multiple component types, Skills co-occur most frequently with every other type, while Commands--Agents is the most common pairing that does not involve Skills.}
Figure~\ref{fig:rq0_component_cooccurrence} shows that Skills appear in nearly every co-occurrence pair. This pattern is unsurprising because Skills are the most prevalent component type in the marketplace, increasing the likelihood that they co-occur with other types. In contrast, the frequent Commands-Agents pairing cannot be explained by prevalence alone, as neither component type dominates the marketplace. This pattern suggests that developers intentionally combine the two. We therefore examine in RQ3 (Section~\ref{sec:rq3}) whether the Commands--Agents pairing reflects functional coupling, where changes to one component require corresponding changes to the other.

\begin{figure}[t]
  \centering
  \includegraphics[width=0.8\linewidth]{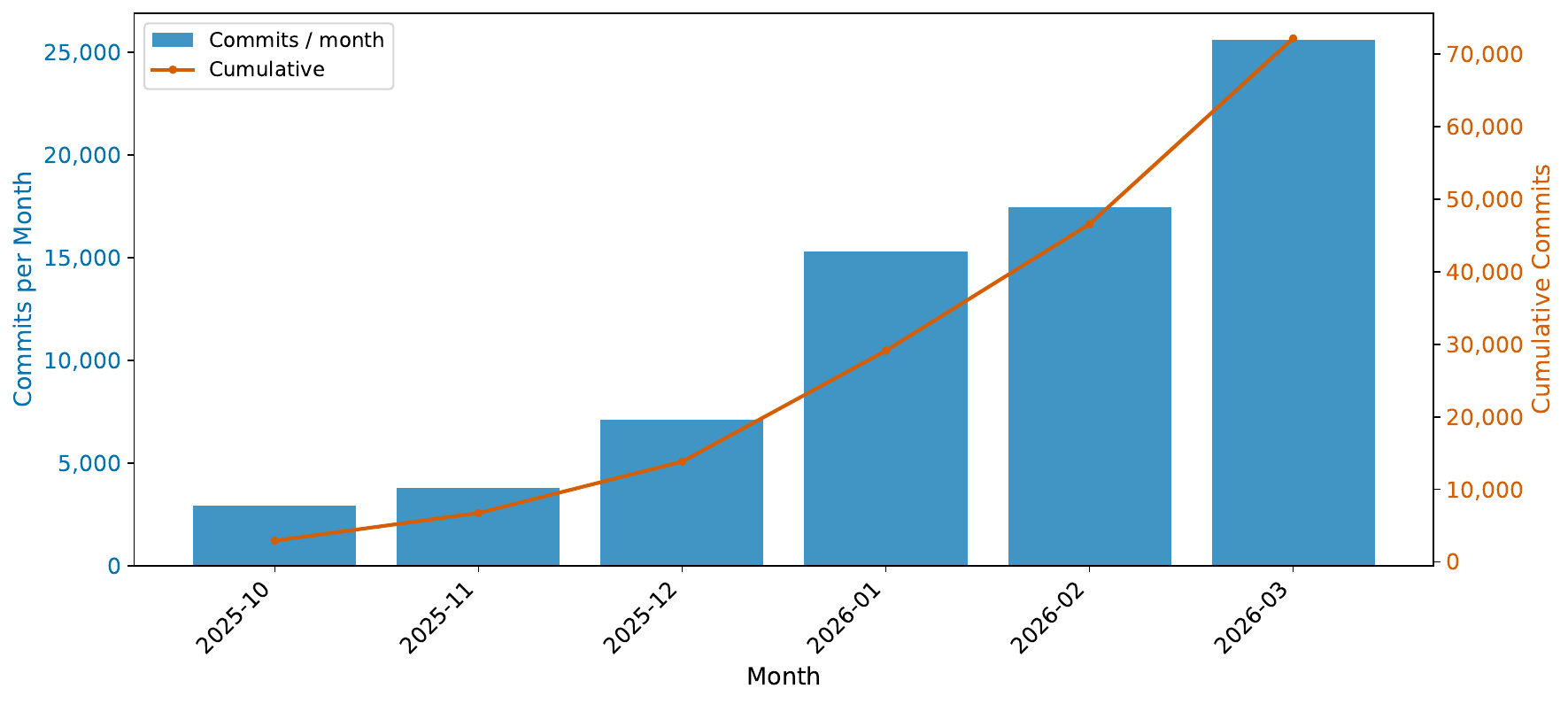}
  \caption{Plugin-touching commits per month (bars) and cumulative
           total (line) since the official plugin launch
           ($n = 72{,}191$. October~2025 through date cutoff at March~2026).
           Each month sets a new high, with no plateau by the
           data cutoff.}
  \label{fig:rq1_monthly_commits}
\end{figure}

\textbf{Plugin-touching commit activity grew 8.8$\times$ between the October 2025 launch and March 2026, rising from 2,923 commits in October 2025 to 25,618 in March 2026, with each month setting a new record.}
Figure~\ref{fig:rq1_monthly_commits} shows monthly commit counts and the cumulative total across 72,191 plugin-touching commits from the official launch through the end of March 2026.
The monthly commit count rose in every month of this period, and a Mann-Kendall trend test confirms that the upward trend is statistically significant with a p-value of $0.003$.
Our Sen's slope test also estimates the month-to-month commit growth at roughly 4,550 additional commits per month. 
The consistent month-over-month growth with no plateau by the data cutoff indicates that agent plugin development activity is still accelerating, raising the question of how that growing effort is organized, which RQ2 addresses by classifying all 77,773 plugin-touching commits by type and authorship.

\begin{summary}{Takeaway (RQ1)}
\begin{itemize}[leftmargin=*]
  \item Software Engineering plugins account for 61.3\% of all plugins, a concentration that mirrors findings for the skills.sh agent skills marketplace~\citep{ling2026agentskills}.
  \item 34.4\% of plugins combine two or more types, with Commands-Agents being the most common pairing.
  \item Plugin-touching commit activity grew 8.8$\times$ in six months with no sign of plateau, indicating the marketplace remains in active growth.
\end{itemize}
\end{summary}

\section{RQ2: \titlecap{\rqtwo}}
\label{sec:rq2}
\subsection{Motivation}

Prior work~\citep{zeng2025ccs} characterizes how developers build and maintain traditional software by studying the kinds of changes they commit, e.g., feature work, bug fix, and maintenance. Our findings from RQ1 show that agent plugin repositories keep accumulating commits after their initial release, and that these repositories consist primarily of natural language instructions alongside scripts and configuration files, rather than program code. Since the artifacts under change differ from program code, the development and maintenance activities behind plugin commits may also differ from those reported for traditional software. In addition, AI coding agents are now widely used in real-world software development~\citep{li2026aidev, li2025rise}, and agent plugin developers are especially likely to delegate work to the agents, raising the question of how much of this development and maintenance work is AI-assisted.

\subsection{Approach}

As software development grows increasingly in complexity, developers adopt commit categorization systems, with the Conventional Commits Specification (CCS) being one of the most widely used conventions~\cite{zeng2025ccs}, to enhance readability for both humans and automated tools.
We therefore adopt the CCS classification system and analyze the commits in agent plugin repositories.
We first apply a two-stage pipeline to assign each commit a label under the CCS classification system to establish a baseline for our study. 
We then conduct a manual study to comprehend the actual development and maintenance behaviors by analyzing the developers' intentions on a subset of commits carrying CCS identifiers.
We then re-classify the commits with an LLM using rubrics summarized from the manual study step to identify CCS types whose meaning in plugin repositories shifts from their original definition in traditional software engineering.

\textbf{Classify commit types under CCS categories.}
We apply a two-stage CCS pipeline following \citet{li2025rise} on all 77,773 plugin-touching commits, assigning each commit exactly one CCS type.
Stage~1 matches commits that already carry a CCS-style prefix (e.g., \texttt{feat:} or \texttt{fix:}) via regular expressions, classifying 51,141 commits (65.7\%) directly.
Stage~2 prompts GPT-5-mini with all 11 CCS definitions plus an \textit{other} category for commits that do not fit any standard type, obtaining a predicted type and confidence score for each of the remaining 26,632 commits (Listing~\ref{lst:ccs_prompt}).

We validate each stage on a separate 385-commit random sample (95\% confidence and a 5\% margin of error~\citep{emse22sampling}), manually labeled by a single rater reading the commit message alongside the changed files.
We apply the Cohen's Kappa with thresholds for interpreting the strength of agreements in \citet{landis1977measurement} and find that Stage~1 reaches 92\% agreement ($\kappa = 0.903$) while Stage~2 reaches 76.4\% ($\kappa = 0.707$), indicating above substantial agreement~\citep{emse22sampling}.
To compare against conventional OSS, we plot our distribution against \citet{zeng2025ccs}, who report CCS type distributions across 88,704 commits from 116 source-code OSS repositories.
To examine how commit effort is distributed across plugin components, we build a heatmap in which each cell shows the percentage of that component type's commits carrying each CCS type, enabling direct comparison across components regardless of their volume.

\textbf{Analyze commit behavior through manual study.}
To understand the meaning of the commits and how they are internally different from traditional software commits, we draw a stratified random sample of 700 commits from the full 77,773 plugin-touching commits (Section~\ref{sec:extraction}), treating each CCS type as a stratum at 95\% confidence and 5\% margin with a minimum of 50 commits per stratum~\citep{emse22sampling}.
We follow an open coding method~\cite{lune2017qualitative} to inspect and classify the sampled commits.
The first two authors first jointly inspect a subset of sampled commits from each CCS type and construct an initial set of categories based on the commit content and the agent plugin context.
The two authors then independently classify the remaining commits based on the initial categories.
When they encounter commits that do not fall into any existing category, the two authors discuss and introduce new categories as needed.
Disagreements are also resolved through discussion until both authors agree on all classification results.
At the end, the two authors inspect the whole set of 700 plugin-touching commits and identify their actual CCS categories (discussed in Section~\ref{sec:rq2_results}), reaching substantial agreement (Cohen's $\kappa = 0.671$).

\textbf{Re-classify commit types whose labels no longer match their function.}
We first identify the CCS types whose meaning in plugin repositories diverges from their traditional software engineering definition, based on the 700 manually inspected plugin commits.
We then re-classify all commits that belong to the CCS types that show definition shifts into appropriate CCS types.
Unlike the \textit{Classify commit types under CCS categories} step, which relies on the developers' intent stated in the commit message, the re-classification in this step reads the actual content diff to discern the commit behavior, the same way the 700 commits were labeled during manual study. 
We summarize the findings from the manual study into per-type rubrics that define CCS categories through development and maintenance behavior in agent plugin scenario (Listing~\ref{lst:relabel_prompt} in Appendix~\ref{app:relabel_prompt}) and leverage GPT-5-mini, the same model used during commit classification under CCS categories, to automate the labeling at scale.
We validate the rubric on the doc, ci, style commit subset of the double-labeled sample, reaching 83\% agreement with the two raters' consensus ($\kappa = 0.65$), comparable to the $\kappa = 0.70$ agreement between the raters themselves.

\textbf{Detect agent contributions.}
To understand what kind of tasks developers delegate to agents, we apply the commit-level heuristics of \citet{robbes2026promises} for detecting the presence of coding agents in software repositories to our 77,773 plugin-touching commits (Section~\ref{sec:extraction}), covering all known agents listed in \citet{robbes2026promises}. For each commit, we scan the message body for co-author trailers (e.g., \texttt{Co-Authored-By: Claude Sonnet 4.5 <noreply@anthropic.com>}) and agent email addresses, and the author field for agent name patterns, all drawn from their catalogue.

\subsection{Results}
\label{sec:rq2_results}

\begin{table}[t]
\centering
\footnotesize
\caption{Development and maintenance behavior observed in a stratified sample of 700 plugin-touching
         commits, one stratum per CCS type.
         Sample sizes follow stratified random sampling~\cite{emse22sampling}:
         $n$\,=\,137 for \textit{feat}, 83 for \textit{fix}, 80 for \textit{chore},
         and 50 per remaining type.
         \textit{fix} and \textit{revert} allow multiple labels per commit so their percentages may not sum to 100\%. 
         \textbf{Bold} indicates the change type within each CCS label that departs most from its conventional software engineering counterpart. 
         \ul{Underline} indicates the sub-categories whose meaning differs from the conventional meaning of their CCS label.
         }
\label{tab:rq2_subcategory}
\begin{tabular}{lp{3.3cm}p{7.7cm}r}
\toprule
\textbf{CCS Type} & \textbf{Change Type} & \textbf{Description} & \textbf{N (\%)} \\
\midrule
\textit{feat} ($n$=137)
& Plugin or component addition
& New \texttt{plugin.json}, marketplace entry, or major component added
& 66 (48.2\%) \\
& \textbf{Component update}
& \textbf{Existing instruction using natural language or config files extended}
& \textbf{54 (39.4\%)} \\
& Incidental touch
& Plugin file changed as a side-effect of non-plugin work
& 17 (12.4\%) \\
\midrule
\textit{fix} ($n$=83)
& \textbf{Instruction/content fix}
& \textbf{Wrong behavioral text corrected in skill, command, or agent files}
& \textbf{41 (49.4\%)} \\
& Hook/command script fix
& Bugs in executable plugin scripts or \texttt{hooks.json}
& 27 (32.5\%) \\
& Plugin manifest fix
& Wrong metadata in \texttt{plugin.json} or \texttt{marketplace.json}
& 12 (14.5\%) \\
& Path fix
& Wrong file paths corrected alongside another fix
& 7 (8.4\%) \\
& Incidental touch
& Plugin file changed as a side-effect of non-plugin work
& 5 (6.0\%) \\
\midrule
\textit{chore} ($n$=80)
& Plugin version bump
& Version field incremented in \texttt{plugin.json} or \texttt{marketplace.json}
& 51 (63.8\%) \\
& Incidental touch
& Plugin file changed as a side-effect of non-plugin work
& 19 (23.8\%) \\
& Manifest/registry update
& Non-version metadata changes (descriptions, skill paths, authors)
& 10 (12.5\%) \\
\midrule
\textit{docs} ($n$=50)
& \ul{\textbf{AI instruction update}}
& \ul{\textbf{\texttt{SKILL.md} or \texttt{agents/*.md} updated, read at runtime by the LLM, developer used \texttt{docs}}}
& \ul{\textbf{37 (74.0\%)}} \\
& Human doc update
& README, CHANGELOG, or release notes; no runtime effect on the AI
& 13 (26.0\%) \\
\midrule
\textit{refactor} ($n$=50)
& \textbf{Instruction/content refactor}
& \textbf{AI-facing Markdown restructured without intended behavior change}
& \textbf{25 (50.0\%)} \\
& File/directory restructuring
& Files or dirs moved or renamed to match schema conventions
& 12 (24.0\%) \\
& Incidental touch
& Plugin file changed as a side-effect of non-plugin work
& 7 (14.0\%) \\
& Component restructuring
& Plugin components extracted, split, merged, or converted
& 6 (12.0\%) \\
\midrule
\textit{test} ($n$=50)
& Incidental touch
& Plugin file changed as a side-effect of non-plugin work
& 24 (48.0\%) \\
& Plugin test file
& pytest, shell, or integration tests for executable plugin scripts
& 14 (28.0\%) \\
& \textbf{AI behavior test file}
& \textbf{Eval cases or pressure tests for model routing and behavior}
& \textbf{7 (14.0\%)} \\
& \textbf{Agent/skill test instruction}
& \textbf{\texttt{SKILL.md} or agent file encoding testing methodology in prose}
& \textbf{5 (10.0\%)} \\
\midrule
\textit{style} ($n$=50)
& \ul{\textbf{Instruction style}}
& \ul{\textbf{Wording, precision, or formatting of AI-facing instruction files}}
& \ul{\textbf{18 (36.0\%)}} \\
& Code style
& Formatting or linting of executable scripts and config JSON
& 16 (32.0\%) \\
& Incidental touch
& Plugin file changed as a side-effect of non-plugin work
& 13 (26.0\%) \\
& \ul{\textbf{AI output style}}
& \ul{\textbf{Verbatim text the AI is instructed to produce reworded}}
& \ul{\textbf{3 (6.0\%)}} \\
\midrule
\textit{perf} ($n$=50)
& \textbf{AI execution management}
& \textbf{Model tier assignment, orchestration, hook tuning, routing accuracy}
& \textbf{24 (48.0\%)} \\
& \textbf{Prompt optimization}
& \textbf{Instruction rewrites to reduce token cost or enable prompt caching}
& \textbf{15 (30.0\%)} \\
& Traditional performance
& Runtime improvements, algorithmic changes, or install/download speed
& 11 (22.0\%) \\
\midrule
\textit{revert} ($n$=50)
& \textbf{Instruction/content reversal}
& \textbf{Original commit primarily changed AI-facing \texttt{.md} files}
& \textbf{25 (50.0\%)} \\
& Config/structure reversal
& Original commit primarily changed plugin config or hooks files
& 16 (32.0\%) \\
& Incidental touch
& Plugin file changed as a side-effect of non-plugin work
& 6 (12.0\%) \\
& Code/script reversal
& Original commit primarily changed executable plugin scripts
& 3 (6.0\%) \\
& Version rollback          [secondary]
& Version number decremented alongside a primary revert label
& 10 (20.0\%) \\
\midrule
\textit{ci} ($n$=50)
& CI/CD workflow management
& GitHub Actions workflows, release pipelines, or validation scripts
& 35 (70.0\%) \\
& \ul{\textbf{CI-domain instruction update}}
& \ul{\textbf{SKILL.md or command files updated inside a plugin that gives Claude CI/CD expertise, \texttt{ci} reflects the plugin's domain, not a pipeline change}}
& \ul{\textbf{14 (28.0\%)}} \\
& Incidental touch
& Plugin file changed as a side-effect of non-plugin work
& 1 (2.0\%) \\
\midrule
\textit{build} ($n$=50)
& Build-label mismatch
& Commit labeled \textit{build} unrelated to actual build activity
& 46 (92.0\%) \\
& Dependency update
& npm bumps, CDN URL changes, or build toolchain updates
& 4 (8.0\%) \\
\bottomrule
\end{tabular}
\end{table}

\textbf{In plugin repositories, \textit{docs}, \textit{perf}, \textit{style}, and \textit{refactor} describe activities that differ from traditional software engineering interpretations.}
Table~\ref{tab:rq2_subcategory} presents the full content category breakdown across the 700-commit sample.
74\% of \textit{docs} commits modify instruction files Claude reads at inference time rather than human-readable documentation, making them functionally equivalent to \textit{feat} or \textit{fix} in traditional SE. 
Whether the developer wrote \texttt{docs:} explicitly or the pipeline inferred it from the Markdown artifact, the label reflects the form of the change, not its function.
\textit{perf} targets model tier assignment, orchestration overhead, and prompt verbosity rather than algorithmic complexity or runtime throughput.
\textit{style} extends into instruction wording precision, where a phrasing change shifts model behavior at runtime rather than affecting code formatting.
\textit{refactor} is dominated by rewording AI-facing instruction text without changing intended behavior and by schema realignment, rather than the code-quality concerns the label carries in conventional repositories.
These four types diverge from their conventional meaning in two ways. \textit{perf} and \textit{refactor} commits hold their CCS labels while referring to natural language AI instructions rather than programming code, whereas \textit{docs} and \textit{style} commits often belong to a different CCS type, indicating a gap that we quantify below.

\textbf{\textit{feat}, \textit{fix}, \textit{chore}, and \textit{ci} retain their conventional meaning in plugin repositories, with \textit{feat} and \textit{fix} shifting from source code to natural-language instruction files.}
As shown in Table~\ref{tab:rq2_subcategory}, \textit{feat} commits frequently introduce new plugins, add components, or extend existing instruction content.
Where conventional feature development centers on implementing code, feature development in plugin repositories often consists of specifying new behavior in natural language.
About half (49.4\%) of \textit{fix} commits correct wrong behavioral text in instruction files rather than code bugs.
Wrong tool references, incorrect workflow steps, and flawed examples in \texttt{SKILL.md} or agent files are the primary defect category in plugin repositories.
In contrast, \textit{chore} and \textit{ci} remain closer to their conventional meaning, with the majority of commits in each type performing traditional maintenance tasks.
\textit{Chore} commits follow traditional release mechanics on plugin-specific artifacts. 63.8\% are version increments in \texttt{plugin.json} or \texttt{marketplace.json}, while 12.5\% are non-version manifest updates registering new skill paths or updating marketplace registry entries.
70.0\% of \textit{ci} commits manage conventional GitHub Actions workflows and release pipelines, while the remainder come from developers of CI-analysis plugins who write \texttt{ci} for instruction updates because their plugin domain is CI tooling, not because any pipeline changed.

\textbf{Instruction Markdown files in plugin repositories are the shared target of \textit{feat}, \textit{fix}, \textit{style}, and \textit{refactor} commits, with developer intent as the only criterion distinguishing them.}
Table~\ref{tab:rq2_subcategory} shows the sub-category descriptions that separate each type across the 700-commit sample.
In conventional repositories, these four types map to structurally different changes.
In plugin repositories, all four can consist of the same text edit to the same instruction file, leaving developer intent as the only criterion.
The boundary between them is inherently ambiguous, as the same diff can qualify as any of the four depending on what the developer intended.

\begin{figure}[t]
  \centering
  \includegraphics[width=0.85\linewidth]{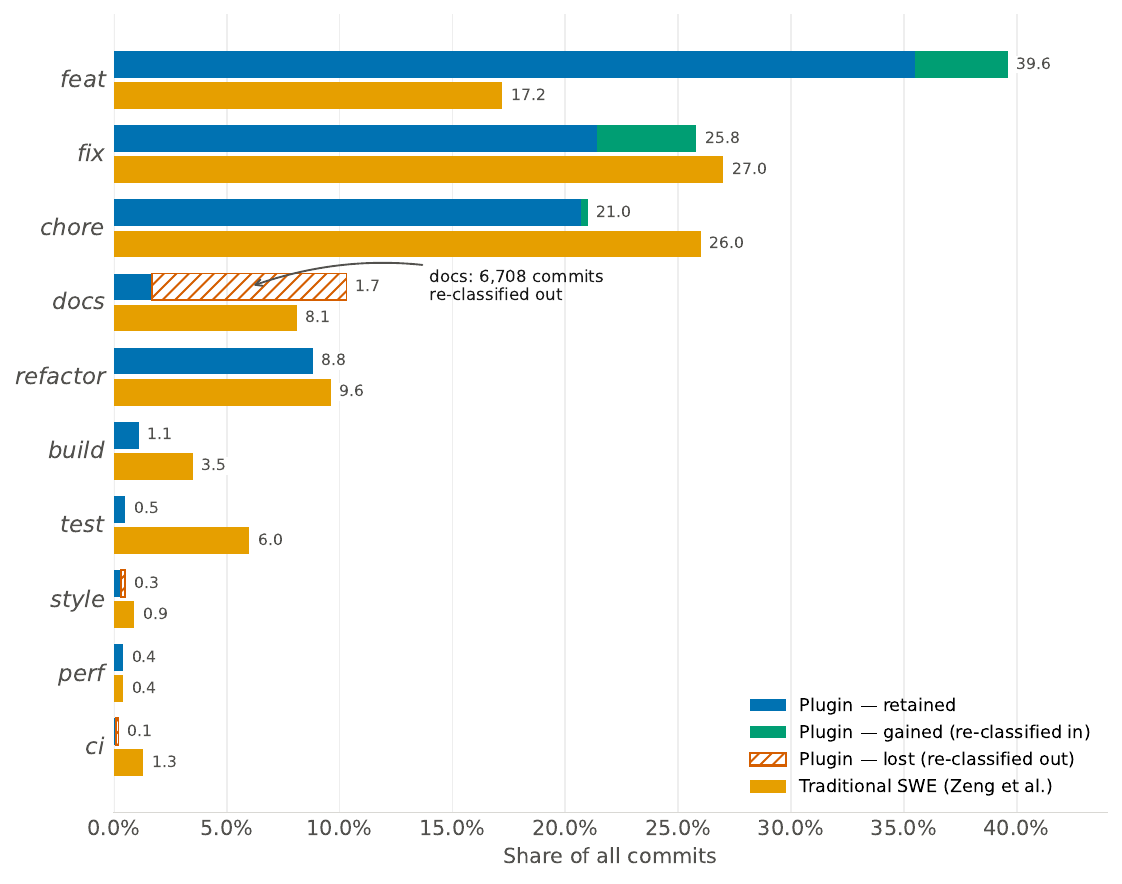}
  \caption{CCS distribution of plugin commits after re-classifying \textit{docs},
  \textit{ci}, and \textit{style} by the function of each commit's diff (blue),
  with the commits each type gains (green) or loses (hatched vermilion) under
  re-classification, against the traditional-SWE baseline (orange,
  \citet{zeng2025ccs}). \textit{docs} collapses from 10.3\% to 1.7\% as its
  commits move to \textit{feat}, \textit{fix}, and \textit{chore}.}
  \label{fig:rq2_delta}
\end{figure}

\textbf{The \textit{docs} share drops from 10.3\% to 1.7\% after re-classification, as 80\% of commits initially labeled \textit{docs} are reassigned to \textit{feat} or \textit{fix}.}
Figure~\ref{fig:rq2_delta} plots the re-classified distribution against the traditional-SWE baseline, where the hatched segment on the \textit{docs} bar marks the 6,709 commits re-classified out of \textit{docs} and the green segments on \textit{feat} and \textit{fix} mark the commits they absorb.
Of the 8,007 commits labeled \textit{docs}, only 16\% remain human documentation, and the rest belong to \textit{fix} (3,353), \textit{feat} (3,047), or \textit{chore} (279) by the work they perform.
More of them are \textit{fix} than \textit{feat}, so much of the documentation activity in plugin repositories corrects the instructions the agent runs rather than adding new ones.
\textit{ci} and \textit{style} shift far less, with 61\% and 62\% keeping their original type, so among the three only \textit{docs} is re-classified at scale.
The re-classification also reverses the comparison with conventional OSS: the raw \textit{docs} share (10.3\%) sits above the 8.1\% reported by \citet{zeng2025ccs}, but by the work these commits perform it falls to 1.7\%, well below that baseline.

\begin{figure}[t]
  \centering
  \includegraphics[width=\linewidth]{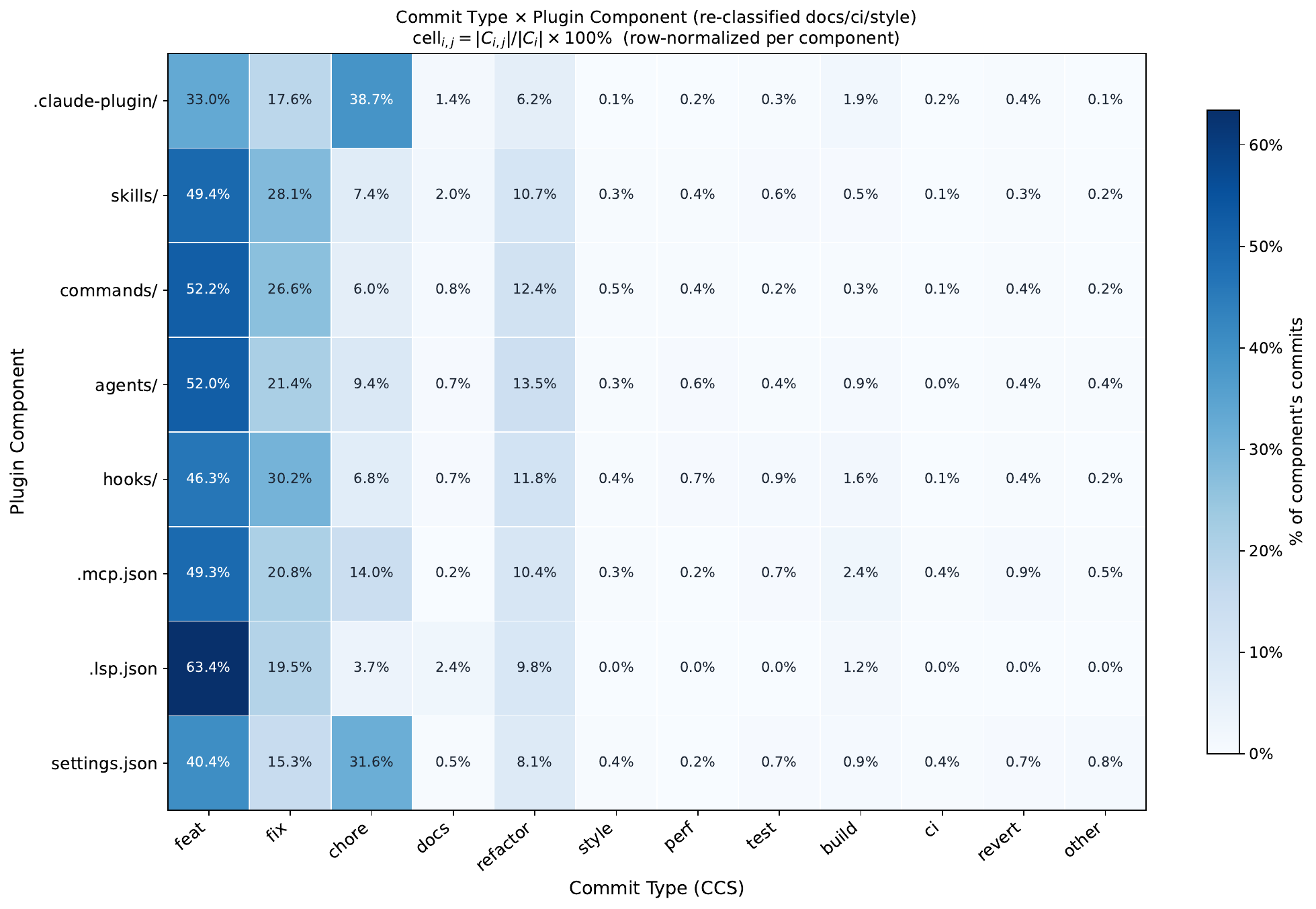}
          \caption{CCS type distribution across plugin components. Each cell shows the percentage of that component's commits carrying a given CCS type, enabling direct comparison across components regardless of their volume.}
  \label{fig:rq2_ccs_heatmap}
\end{figure}

\textbf{\textit{feat} dominates plugin commits at 39.6\%, more than double its 17.2\% share in conventional OSS, while \textit{chore} at 21.0\% shows that developers actively return to maintain their plugins after initial creation.}
Figure~\ref{fig:rq2_delta} shows \textit{feat}, \textit{fix}, and \textit{chore} together account for 86.5\%~(67,271 out of 77,773 commits) of all classified commits, with \textit{feat} alone at 39.6\%~(30,820 out of 77,773).
The \textit{feat} dominance reflects the low-friction authoring model of Markdown-based plugins, where adding a capability requires writing natural language instructions rather than programming language code.
While \textit{feat} reflects the ease of initial plugin authoring, the substantial \textit{chore} share (21.0\%), primarily due to version increments and manifest updates, shows that developers do return to maintain their plugins rather than abandoning them after initial creation.
\textit{fix} ranks second at 25.8\%, below the 27\% it holds in conventional OSS~\citep{zeng2025ccs}, reflecting the corrective maintenance burden of programming language artifacts.
\textit{build} and \textit{test} are also substantially less common in plugin repositories (1.1\% and 0.5\%, respectively) than in conventional OSS, since natural-language artifacts require no compilation pipeline or dedicated test suite, though the Zeng et al.\ repositories are substantially older, which may amplify their fix share (Section~\ref{sec:threats}).

\textbf{\textit{feat} is the dominant commit type across nearly all plugin component types, with \texttt{.claude-plugin/} as the sole exception where \textit{chore} leads.}
Figure~\ref{fig:rq2_ccs_heatmap} shows how commit intent shifts across each component type.
Among behavioral components, \textit{feat} accounts for 46.3--52.2\% of commits in \texttt{skills/}, \texttt{commands/}, \texttt{agents/}, and \texttt{hooks/}.
In contrast, \textit{chore} dominates at 38.7\% versus \textit{feat} at 33.0\% for \texttt{.claude-plugin/}.
This exception reflects the nature of \texttt{.claude-plugin/}, which stores version manifests, schema declarations, and marketplace registrations that accumulate recurring configuration updates rather than new behavioral additions.
These updates account for much of the overall \textit{chore} share shown in Figure~\ref{fig:rq2_delta}.

\begin{table}[t]
\centering
\caption{Agent contributions and Claude co-authorship across 77,773 plugin-touching commits.}
\label{tab:rq2_agent_coauth}
\begin{subtable}[t]{0.53\linewidth}
    \centering
    \caption{Multi-agent detection results following \citet{robbes2026promises},
             covering all known agents listed in their study.
             The heuristics were applied to $n=77{,}773$ commits.
             Agents with zero commit-level matches are omitted.}
    \label{tab:rq2_multiagent}
    \normalsize
    \begin{tabular}{llrr}
    \toprule
    \textbf{Agent} & \textbf{Heuristics} & \textbf{\# Commits} & \textbf{\% } \\
    \midrule
    Claude      & trailer, email, author & 27,159 & 34.9   \\
    Copilot     & trailer, email, author &    331 & 0.4    \\
    Cursor      & trailer, email, author &     80 & 0.1    \\
    Amp         & trailer                &     65 & $<$0.1 \\
    Codex       & trailer, email, author &     24 & $<$0.1 \\
    Gemini      & trailer, name pattern  &     17 & $<$0.1 \\
    Devin       & trailer, author        &     12 & $<$0.1 \\
    CodeRabbit  & trailer                &     10 & $<$0.1 \\
    Sourcery    & trailer                &      5 & $<$0.1 \\
    Jules       & trailer                &      4 & $<$0.1 \\
    Opencode    & trailer                &      3 & $<$0.1 \\
    Aider       & trailer, email, author &      1 & $<$0.1 \\
    ChatGPT     & trailer                &      1 & $<$0.1 \\
    QwenCoder   & trailer                &      1 & $<$0.1 \\
    \midrule
    \textbf{Any agent} &                & \textbf{27,636} & \textbf{35.5} \\
    \bottomrule
    \end{tabular}
\end{subtable}%
\hfill
\begin{subtable}[t]{0.45\linewidth}
    \centering
    \caption{Claude co-authorship rate by CCS type following
             the heuristics of \citet{robbes2026promises}, after
             re-classifying \textit{docs}, \textit{ci}, and \textit{style}
             commits (Section~\ref{sec:rq2_results}).}
    \label{tab:rq2_ccs_distribution_corrected}
    \small
    \begin{tabular}{lrrr}
    \toprule
    \textbf{CCS} & \textbf{\# Commits} & \textbf{\# Claude} & \textbf{\% Claude} \\
    \midrule
    feat      & 30,820 & 12,234 & 39.7 \\
    fix       & 20,089 &  7,723 & 38.4 \\
    chore     & 16,362 &  3,798 & 23.2 \\
    refactor  &  6,872 &  2,438 & 35.5 \\
    docs      &  1,299 &    405 & 31.2 \\
    build     &    854 &    161 & 18.9 \\
    test      &    425 &    132 & 31.1 \\
    perf      &    287 &    115 & 40.1 \\
    revert    &    265 &     43 & 16.2 \\
    style     &    248 &     51 & 20.6 \\
    other     &    139 &     24 & 17.3 \\
    ci        &    113 &     35 & 31.0 \\
    \midrule
    \textbf{Total} & \textbf{77,773} & \textbf{27,159} & \textbf{34.9} \\
    \bottomrule
    \end{tabular}
\end{subtable}

\end{table}

\textbf{35.5\% of plugin commits are co-authored by an AI coding agent, with Claude accounting for 34.9\% and dominating AI-assisted plugin development and AI contribution most commonly seen in \textit{perf} commits (40.1\%).}
Table~\ref{tab:rq2_agent_coauth} reports detection results for all agents in \citet{robbes2026promises}'s catalogue alongside Claude's co-authorship rate, broken down by commit type.
Overall, 35.5\% of plugin commits match at least one coding agent.
Of these, 77 commits (0.1\%) match more than one agent.
All other coding agents combined contribute less than 0.5\%, making the marketplace effectively a single-agent environment. This partly reflects the dataset. The plugins studied are built for Claude Code, where Claude is the natural co-author.
Co-authorship ranges from 40.1\% for \textit{perf} commits to 16.2\% for \textit{revert} commits, a gap of 23.9 percentage points, suggesting that developers use Claude depending on task type rather than uniformly across all development tasks.

\begin{summary}{Takeaway (RQ2)}
\begin{itemize}[leftmargin=*]
  \item A manual analysis of 700 commits reveals that four CCS types carry different meanings in plugin repositories: \textit{docs} commits predominantly update runtime AI behavior, \textit{perf} targets prompt efficiency rather than algorithmic speed, \textit{style} extends to instruction wording, and \textit{refactor} is dominated by schema realignment.
  \item Plugin development is predominantly feature-driven. \textit{feat}, \textit{fix}, and \textit{chore} account for 86.5\% of all commits, with \textit{feat} at 39.6\% reflecting the ease of adding capabilities through natural-language instruction files.
  \item AI coding agents co-author 35.5\% of all plugin commits, with co-authorship ranging from 40.1\% for \textit{perf} to 16.2\% for \textit{revert} commits.
\end{itemize}
\end{summary}

\section{RQ3: \titlecap{\rqthree}}
\label{sec:rq3}
\subsection{Motivation}
Prior work reports that in traditional OSS, infrastructure code and source code co-evolve tightly, which introduces additional efforts for the people responsible for these changes to keep the project compilable, deployable, or executable~\citep{jiang2015coevolution}.
Our findings from previous research questions show that agent plugins frequently combine multiple component types and languages (both natural language and multiple programming languages) within the same repository (RQ1), with significant contribution from AI coding agents (RQ2).
The distinct composition of components, languages, and contributions indicates that the prior findings on co-evolution and coupling on traditional OSS might not transfer well to agent plugin repositories.

Therefore, we investigate whether agent plugins have similar coupling between components, examining potential co-evolutions at two levels: 1) inter-component and 2) intra-component.
Since development is predominantly natural-language driven (RQ2), when scripts and Markdown files change together, classifying what drives that inter-component coupling reveals how maintaining natural-language artifacts differs from maintaining source code.
Intra-component co-evolution operates within individual components themselves, where the same question arises at the file-type level, since skills and hooks mix file types that could co-evolve within a single directory.
Together, the answers offer researchers the first empirical co-evolution measurements for AI-native repositories, where the dependency of interest runs not between two code files but between an implementation script and the natural-language instructions an agent reads at runtime.

\subsection{Approach}

To study how plugin components co-evolve, we first collect merged pull requests (PRs), then conduct analyses using the dataset described in Section~\ref{sec:extraction}.

\textbf{Collect merged pull requests.}
To capture developer-reviewed units of change in which related modifications are bundled together, making cross-component co-change more visible within individual commits~\citep{barrak2021coevolution}, we collect merged pull requests via the GitHub REST API across all 1,926 repositories.
We retrieve closed pull requests and retain only those with a non-null \texttt{merged\_at} timestamp, since only merged pull requests represent changes accepted into the code
base, capping at 500 changed files per pull request to avoid timeouts on large non-plugin changes. 
This yields 112,807 merged pull requests.
For the coupling analyses, we restrict to the 748 repositories with more than ten merged pull requests following \citet{barrak2021coevolution} and to multi-component plugins (those containing at least two distinct component types), since association rules between component types require both types to appear in the data, yielding 10,679 plugin pull requests from 2,592 plugins.

\textbf{Measure cross-component coupling.}
Association rules measure how often two items co-occur in transactions. 
Applied to pull requests, they capture how often two component types change together in the same pull request and whether that co-change is stronger than chance.
To understand whether updating one component forces updates to others, we apply the association rule methodology first introduced for build co-evolution by \citet{mcintosh2011build} and later adopted by \citet{jiang2015coevolution} with merged pull requests as the unit of work.
For each directed component type pair (X$\rightarrow$Y and Y$\rightarrow$X are measured separately since Confidence is asymmetric), we compute Support, Confidence, Lift, and a chi-squared significance test ($\alpha = 0.05$) without an additional multiple-testing correction, consistent with \citet{jiang2015coevolution}:
\begin{equation}
Supp(X \Rightarrow Y) = Supp(Y \Rightarrow X) = P(X \cap Y)
\label{eq:support}
\end{equation}
\begin{equation}
Conf(X \Rightarrow Y) = \frac{P(X \cap Y)}{P(X)}
= \frac{Supp(X \Rightarrow Y)}{P(X)}
\label{eq:confidence}
\end{equation}
\begin{equation}
Lift(X \Rightarrow Y) = \frac{P(X \cap Y)}{P(X)P(Y)}
= \frac{Conf(X \Rightarrow Y)}{P(Y)}
\label{eq:lift}
\end{equation}
\begin{equation}
\chi^2(X \Rightarrow Y)
= n(Lift-1)^2
\frac{Supp \times Conf}{(Conf-Supp)(Lift-Conf)}
\label{eq:chi-square}
\end{equation}
\noindent Support measures how often X and Y change together across all pull requests. Confidence measures how often Y changes given that X already changed in the same pull request. Lift contextualizes this Confidence by accounting for how frequently Y changes overall, with above 1 indicating that Y changes more often with X than expected from their individual frequencies~\citep{barrak2021coevolution}.

\textbf{Measure intra-component coupling.}
To assess whether different file types co-change within individual component directories, we apply the same association rule methodology within \path{skills/} and \path{hooks/}, using file types as the unit of work rather than component types.
Any plugin with a \path{skills/} or \path{hooks/} directory qualifies, since the analysis requires only two distinct file types within a single directory rather than two distinct component types across a plugin, yielding 13,587 intra-skills pull requests from 5,353 plugins and 1,696 intra-hooks pull requests from 1,085 plugins across the same 748 repositories.

\textbf{Filter and sample co-change pull requests.}
To study what happens when scripts and Markdown files change together, we apply two filters to the 13,587 intra-skills pull requests. First, we retain only pull requests where at least one script (.py, .sh, or .ts) and at least one Markdown file co-change within the same \path{skills/} subdirectory, yielding 1,908 pull requests. Second, we restrict to pull requests where all such files are modified rather than newly created, excluding pull requests where the script and its paired Markdown file are introduced together in the same pull request, yielding 323 pull requests.
We draw a stratified sample of 64 pull requests (20\% per stratum) across the six script-type $\times$ Markdown-type combinations (.py, .sh, .ts $\times$ SKILL.md, Other .md), ensuring all co-change combinations are represented when we construct the taxonomy.

\textbf{Classify co-change patterns.}
We read the full script and Markdown diff of each pull request and classify what the developer changed in the script and why the Markdown required corresponding updates. 
Following \citet{dig2006apis}, who distinguish structural changes that preserve a component's behavior from behavioral changes that modify it, we characterize each co-change by the nature of the underlying script change rather than by the file type touched. We do not reuse their object-oriented catalog directly. Instead, we derive five plugin-specific coupling categories through open coding of the sample.
We follow the same open coding process used for the commit content categories (RQ2): the first two authors jointly inspect a small set of pull requests and construct an initial set of categories based on the nature of the underlying script change, then independently classify more pull requests based on the initial categories, discussing and adding a new category as needed when a pull request does not fit an existing one.
The two raters independently coded all 64 pull requests against the resulting scheme, reaching substantial agreement (Cohen's $\kappa = 0.74$), and resolved the 13 disagreements through discussion until every pull request had a single agreed label.

\textbf{Scale the taxonomy with an LLM classifier.}
To extend the taxonomy built from the 64 human-coded pull requests to the remaining 259, we apply an LLM classifier, following the same LLM-assisted scaling approach used for commit classification in RQ2. We prompt a \texttt{gpt-5-mini} classifier to assign one of the five categories to each co-change pull request from its full script and Markdown diff, treating the 64 human-coded pull requests as ground truth.
We validate the classifier's predictions against this ground truth, reaching substantial agreement (Cohen's $\kappa = 0.62$), then apply it to the remaining 259 co-change pull requests.
The classification prompt is provided in Appendix~\ref{app:coupling_prompt}.

\subsection{Results}

\textbf{The inter-component analysis shows that skills co-change with every other component type in at least 43.2\% of pull requests where the other type changes, showing that keeping skills consistent with the rest of a plugin is a recurring effort.}
Table~\ref{tab:rq3_cochange} reports the association rule metrics for all directed component pairs across 10,679 plugin pull requests. Confidence reaches at least 43.2\% in all four component to skills rules, peaking at 56.6\% for agents to skills.
The repository significance column shows the percentage of repositories in which the pair couples significantly within that repository on its own, ranging from 16.9\%--27.9\% across pairs.
Prior studies provide context for the magnitude of these co-change rates. \citet{jiang2015coevolution} report median Confidence values between 3.5\% and 45.8\% for infrastructure-code coupling, while \citet{barrak2021coevolution} report median Confidence values between 27.46\% and 91.91\% for DVC-source coupling at the pull-request level. The 43--57\% component to skills Confidence falls within the range observed for cross-artifact co-evolution in these settings, showing that changes to other plugin components frequently coincide with changes to skills.

Lift (Eq.~\ref{eq:lift}) provides a potential explanation for these high co-change rates reflect the dominance of skills.
Because skills is the most common component type, it changes in the majority of pull requests regardless of what else changes, so every component-to-skills Lift falls below 1 (0.59--0.78).
For example, agents change with skills in 56.6\% of pull requests, a substantial and recurring amount of co-change work, so the co-change is real.
Lift is only 0.78, so this is likely due to the dominance of skills and we cannot attribute it to a coupling specific to agents and skills.
In contrast, agents--commands reaches Lift 1.40 ($p < 0.05$), the only one of the five reported pairs where co-change exceeds chance.
Further analysis reveals that the coupling between agents and commands is bidirectional.
We observe that the agent definitions embed command slash-names in their output text, and commands reference specific agents in their delegation instructions. For example, an agent definition instructs ``Run \texttt{/rune:activate} to retry'', while a command's instructions read ``You MUST use the Task tool to launch 3 reviewer agents in parallel''.
As each component explicitly references the other, renaming either one requires updating all corresponding files that mention it.

\begin{table}[t]
\centering
\caption{Association rule metrics for directed component type pairs
         ($n=10{,}679$ plugin pull requests).
         Pairs with fewer than 100 co-changing pull requests (PRs) are omitted,
         as are pairs whose coupling is not statistically significant ($\chi^2 \leq 3.84$). 
         Repos significance shows the percentage of repositories with
         statistically significant coupling ($\chi^2 > 3.84$).}
\label{tab:rq3_cochange}
\begin{tabular}{lrrrr}
\toprule
\textbf{Rule ($X \rightarrow Y$)} & \textbf{Co-changing PRs}
  & \textbf{Conf} & \textbf{Lift} & \textbf{Repos significance} \\
\midrule
skills $\rightarrow$ commands  & 1,162 & 14.9\% & 0.64 & 23.4\% \\
commands $\rightarrow$ skills  & 1,162 & 46.5\% & 0.64 & 23.4\% \\
\addlinespace
skills $\rightarrow$ agents    & 1,176 & 15.1\% & 0.78 & 25.8\% \\
agents $\rightarrow$ skills    & 1,176 & 56.6\% & 0.78 & 25.8\% \\
\addlinespace
skills $\rightarrow$ hooks     &   678 &  8.7\% & 0.59 & 27.9\% \\
hooks $\rightarrow$ skills     &   678 & 43.2\% & 0.59 & 27.9\% \\
\addlinespace
agents $\rightarrow$ commands  &   679 & 32.7\% & 1.40 & 25.0\% \\
commands $\rightarrow$ agents  &   679 & 27.2\% & 1.40 & 25.0\% \\
\addlinespace
skills $\rightarrow$ MCP       &   146 &  1.9\% & 0.65 & 16.9\% \\
MCP $\rightarrow$ skills       &   146 & 47.2\% & 0.65 & 16.9\% \\
\bottomrule
\end{tabular}
\end{table}

\begin{table}[t]
\centering
\caption{Intra-skills file-type co-change association rules
         ($n=13{,}587$ pull requests from 5,353 plugins).
          Script-to-script pairs and pairs with fewer than 200 co-changing pull requests are omitted, as are pairs whose coupling is not statistically significant ($\chi^2 \leq 3.84$).
         Other .md corresponds to supporting files~\citep{claudecode_skills}.
         Repos significance\ shows the percentage of repositories with significant
         coupling ($\chi^2 > 3.84$).}
\label{tab:rq3_intra}
\small
\begin{tabular}{lrrrr}
\toprule
\textbf{Rule ($X \rightarrow Y$)} & \textbf{Co-changing PRs}
  & \textbf{Conf} & \textbf{Lift} & \textbf{Repos significance} \\
\midrule
.py $\rightarrow$ README.md       & 201 & 19.1\% & 4.14 & 33.3\% \\
README.md $\rightarrow$ .py       & 201 & 32.1\% & 4.14 & 33.3\% \\
\addlinespace
.json $\rightarrow$ README.md     & 203 & 17.5\% & 3.80 & 48.6\% \\
README.md $\rightarrow$ .json     & 203 & 32.4\% & 3.80 & 48.6\% \\
\addlinespace
.py $\rightarrow$ Other .md       & 670 & 63.6\% & 1.58 & 16.0\% \\
Other .md $\rightarrow$ .py       & 670 & 12.2\% & 1.58 & 16.0\% \\
\addlinespace
.sh $\rightarrow$ Other .md       & 289 & 55.0\% & 1.37 & 23.1\% \\
Other .md $\rightarrow$ .sh       & 289 &  5.3\% & 1.37 & 23.1\% \\
\addlinespace
.json $\rightarrow$ Other .md     & 548 & 47.3\% & 1.17 & 12.4\% \\
Other .md $\rightarrow$ .json     & 548 & 10.0\% & 1.17 & 12.4\% \\
\addlinespace
SKILL.md $\rightarrow$ .json      & 694 &  6.2\% & 0.73 & 11.7\% \\
.json $\rightarrow$ SKILL.md      & 694 & 59.9\% & 0.73 & 11.7\% \\
\addlinespace
README.md $\rightarrow$ Other .md & 360 & 57.4\% & 1.43 & 26.1\% \\
Other .md $\rightarrow$ README.md & 360 &  6.6\% & 1.43 & 26.1\% \\
\addlinespace
Other .md $\rightarrow$ SKILL.md  & 4,086 & 74.6\% & 0.91 & 15.8\% \\
SKILL.md $\rightarrow$ Other .md  & 4,086 & 36.5\% & 0.91 & 15.8\% \\
\bottomrule
\end{tabular}
\end{table}

\textbf{The intra-component analysis shows that, scripts and Markdown files co-change in at least 55\% of pull requests within a single component, showing above-chance coupling within \path{skills/}.}
Table~\ref{tab:rq3_intra} reports the intra-component association rules across 13,587 skill pull requests and 1,696 hook pull requests.
Confidence ranges from 55--64\% for script--Markdown pairs within \path{skills/} (all $p < 0.05$).
\path{SKILL.md} appears in 82.5\% of all skill pull requests regardless of what else changes, suggesting developers consistently update it whenever a skill changes.
In contrast, hooks function as single-language isolated scripts with no cross-file coupling.
Unlike the inter-component results, Lift here exceeds 1 for all script--Markdown pairs (1.37--1.58), confirming that scripts and Markdown files co-change more than chance alone would predict within a single skill directory.

\begin{table}[t]
\caption{Script--Markdown co-change categories. The scheme was built by open coding a sample of
64 pull requests (two raters, Cohen's $\kappa=0.74$) and applied to the remaining 259 by a
\texttt{gpt-5-mini} classifier (validated at $\kappa=0.62$ against the 64). 
The first four categories
are coupling mechanisms, and the remaining No coupling category shows no correlation between script and Markdown file changes.}
\label{tab:intra-cochange}
\centering\small
\begin{tabular}{lp{9.2cm}rr}
\toprule
\textbf{Category} & \textbf{Description} & \textbf{Human} & \textbf{LLM} \\
                  &                      & \textbf{(n=64)} & \textbf{(n=259)} \\
\midrule
Interface change      & A flag, command, argument, or method is added or removed in the
                        script; \path{SKILL.md} updates the callable surface the agent
                        invokes. & \textbf{17 (27\%)} & 87 (34\%) \\
Internal logic change & The script's behavior or output changes while it is invoked the
                        same way; \path{SKILL.md} updates its description of what the
                        skill does. & \textbf{17 (27\%)} & 68 (26\%) \\
Variable/version sync  & A version, default, or identifier the script embeds changes
                        (model name, environment variable, version string); every
                        Markdown file that quotes it follows. & \textbf{9 (14\%)} & 33 (13\%) \\
Repo re-structuring   & A file, path, or configuration-file format is moved, renamed,
                        or reformatted; \path{SKILL.md} updates the affected paths,
                        filenames, or format examples. & \textbf{7 (11\%)} & 24 (9\%) \\
\midrule
No coupling           & The script and the Markdown both change, but the script change
                        does not drive the Markdown file change. & 14 (22\%) & 47 (18\%) \\
\bottomrule
\end{tabular}
\end{table}

\textbf{50 of the 64 sampled pull requests (78\%) contain at least one functionally coupled script--Markdown co-change.}
Table~\ref{tab:intra-cochange} summarizes the four coupling categories.
Interface changes and internal logic changes are the two most frequent drivers, each covering 17 of the 50 coupled pull requests (34\%). 
When a developer adds or removes a flag, command, or method, or changes what the script computes or outputs, \path{SKILL.md} must be rewritten before the agent can invoke the skill correctly.
Variable and version synchronization (9 pull requests), where a model identifier, environment variable, or version string changes in the script, every Markdown file that quotes it must follow, because these variables exist only in the natural language that configures the skill. In one sampled pull request, a single model-identifier change propagated to 24 \path{SKILL.md} files.
Repo re-structuring (7 pull requests) covers changes where a renamed path, filename, or configuration-file format forces matching edits to every instruction that references it.
The remaining 14 pull requests (22\%) show no coupling. In these pull requests the script and the Markdown file both change, but the script update is not what drives the Markdown edit. The \texttt{gpt-5-mini} classifier, which reaches substantial agreement with the human labels (Cohen's $\kappa = 0.62$), reproduces this distribution almost exactly on the remaining 259 co-change pull requests (Table~\ref{tab:intra-cochange}, LLM column), confirming the taxonomy generalizes beyond the manually coded sample.

\begin{figure}[t]
  \centering
  \includegraphics[width=\linewidth]{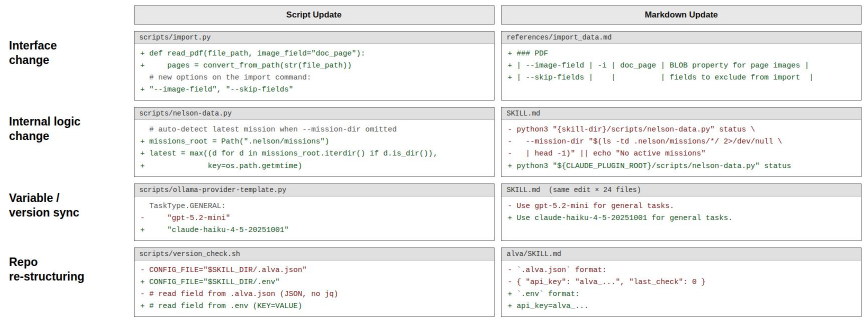}
  \caption{Examples of script and Markdown changes across the coupling categories, selected from real plugin pull requests. For each example, the script update is shown on the left and the related Markdown update on the right.}
  \label{fig:coupling-categories}
\end{figure}

Figure~\ref{fig:coupling-categories} illustrates one representative diff pair per category.
For interface changes, an import script gained two new flags (\texttt{--image-field}, \texttt{--skip-fields}), and \path{SKILL.md} added a section covering both flags and the updated invocation.
For internal logic changes, a status command was extended to auto-detect the latest mission directory, and the \path{SKILL.md} invocation shrank from an 80-character shell pipeline the agent had to compose verbatim to a single plain call.
For variable and version synchronization, a provider template changed one model identifier, and 24 \path{SKILL.md} files were updated in the same pull request because the model name was quoted in every sub-skill's instructions.
For repo re-structuring, a skill switched its configuration file from \texttt{.alva.json} to \texttt{.env}, and \path{SKILL.md} replaced every JSON example with the dotenv format.

\begin{summary}{Takeaway (RQ3)}
\begin{itemize}[leftmargin=*]
  \item The inter-component analysis shows skills co-changing with every other component type in at least 43.2\% of pull requests, though Lift stays below 1 (0.59--0.78), and only agents--commands co-change above chance.
  \item The intra-component analysis shows scripts and Markdown files inside \path{skills/} co-changing above chance (Lift 1.37--1.58), revealing dependencies not visible at the component level.
  \item Manual inspection of 64 sampled co-change pull requests confirms that 78\% are functionally coupled, driven by interface and internal-logic changes that propagate from scripts to their paired instruction files.
\end{itemize}
\end{summary}

\section{Implications}
\label{sec:implications}

\subsection{Implications for Researchers}

\textbf{Researchers should treat AI-facing Markdown as a first-class software artifact whose maintenance requires dedicated analysis methods beyond code-centric metrics.} RQ2 shows that 74\% of \textit{docs} commits in plugin repositories modify instructions read by Claude at runtime rather than human-readable documentation, making them functionally equivalent to \textit{feat} or \textit{fix} commits in traditional SE. 
Code-centric measures such as complexity, clone detection, and test density cannot capture the semantic drift, instruction staleness, or capability coverage of the natural-language artifact that directly controls Claude's behavior at runtime, and tracking maintenance quality in plugin repositories requires new analysis methods designed for natural-language content rather than recalibrated code metrics.
Researchers should develop analyses tailored to the semantic structure of natural-language instruction files, including semantic differencing to detect meaning changes without surface-level edits, instruction linting to identify missing or contradictory directives, and consistency metrics that verify Markdown plugin files remain synchronized with their paired scripts.
Whether commits that touch both plugin component files and non-plugin artifacts such as source code, configuration, and documentation reflect systematic coupling between plugins and their host repositories or coincidental project activity remains an open question, and resolving it would determine whether instruction-aware analyses need to extend beyond plugin boundaries to provide complete coverage.

\textbf{Researchers should not apply commit classifiers trained on conventional OSS directly to AI-native plugin repositories, as commit-type semantics diverge substantially across ecosystems.} Our manual analysis of 700 commits in (RQ2) shows that four CCS types carry substantially different meanings in plugin repositories.
Applying commit classifiers trained on conventional OSS to plugin repositories, or comparing CCS distributions without semantic validation, will produce misleading results, in plugin repositories, \textit{docs} commits most often modify runtime AI behavior, and \textit{perf} commits target prompt efficiency rather than algorithmic speed.
Researchers building commit analysis tools for plugin ecosystems should fine-tune automated classifiers on the semantically validated plugin-specific labels, and researchers studying other AI-native ecosystems should validate whether the same semantic shifts appear before applying conventional CCS taxonomies.
Whether the same semantic drift holds in the emerging plugin marketplace of other AI coding agents, such as GitHub Copilot and OpenAI Codex, where natural-language-first artifacts may produce analogous label shifts, remains untested.

\textbf{Researchers should study human-AI collaboration in AI-native plugin repositories.} Claude co-authors 34.9\% of all plugin-touching commits with measurable variation across types, from 40.1\% in \textit{perf} to 16.2\% in \textit{revert} (RQ2).
These agent plugin marketplaces expose a visible division of labor between humans and coding agents across tens of thousands of real commits, providing large-scale observational data on human-AI collaboration.
Researchers should use this marketplace to study delegation patterns, review dynamics, and how human and agent responsibilities evolve as AI-native projects mature.
Whether the human-AI co-authorship profile shifts over a plugin's lifecycle remains uncharacterized. Tracking whether the high \textit{feat} co-authorship rate of the initial growth phase gives way to increasing \textit{fix} and \textit{perf} co-authorship as a plugin matures would reveal whether AI involvement is a persistent property of the maintenance regime or concentrated in early development.

\textbf{Software maintenance researchers should develop consistency-checking techniques for code--instruction dependencies.} 78\% of Script--Markdown co-changes in our sample are functionally necessary, driven by interface and internal-logic changes, with variable renames cascading across instruction files (RQ3).
Unlike traditional documentation drift, a stale \path{SKILL.md} causes Claude to invoke the skill incorrectly at runtime, making instruction staleness a correctness defect rather than a readability issue with no compiler or type checker to flag the mismatch at development time, analogous to cross-language interface errors where a signature change in one language silently breaks callers in another.
Researchers should build automated techniques that extract callable interfaces from scripts and verify that corresponding Markdown instructions, parameter names, and usage examples remain synchronized.
Automated propagation of interface changes to natural-language files is a tractable and unsolved problem.
The four co-change categories in Section~\ref{sec:rq3} each have a predictable structure detectable from a script diff, a single identifier rename already propagates to dozens of Markdown files in practice, and no existing tool addresses this class of dependency.

\textbf{Researchers studying AI coding agent platforms should plan longitudinal follow-ups of plugin marketplaces.} Commit activity grew 8.8$\times$ between the October 2025 launch and March 2026 with no sign of plateau at the data cutoff (RQ1). The rapid post-launch creation of repositories also produced structural redundancy: 38.7\% of normalized plugin names appear more than once, with the majority tracing to collective aggregator repositories that re-register plugins from other sources, meaning raw plugin counts overstate true supply breadth.
As more developers from diverse professional domains join this expanding marketplace, the current concentration in Software Engineering (61.3\% of all plugins) is likely to evolve, with other fields potentially growing into substantial segments.
Researchers should track how the category distribution evolves as the marketplace grows, whether developers shift focus toward other fields or whether SE concentration deepens further, and whether multi-component plugins transition from a minority (34.4\%) to the dominant architectural pattern.
Whether the Software Engineering concentration and Skills dominance observed here represent a temporary signature of the early growth phase or a persistent structural property of AI coding agent marketplaces remains unknown, and answering it would require longitudinal tracking of this marketplace or cross-platform comparisons at different stages of maturity.

\subsection{Implications for Plugin Developers}

\textbf{Plugin developers who modify Agent or Command definitions should verify that cross-references between the two remain consistent.} Agents--Commands is the only component pair with above-chance co-change coupling across 10,679 plugin pull requests (RQ3).
Some agent definitions embed command slash-names in their output text and some commands embed agent names in their delegation instructions. In these cases, renaming either artifact leaves the other referencing a name that no longer exists, with no static check or compiler to flag the broken reference before Claude attempts the invocation at runtime.
Developers should verify, before merging any rename to an Agent or Command file, that every cross-reference to it in the paired artifact still resolves to a valid name.
Whether the coupling between Agents and Commands grows stronger as a plugin matures, or whether newer component types such as MCP servers become the dominant co-change pair over time, remains an open question.

\textbf{Plugin developers who modify scripts in \path{skills/} directories should verify whether a corresponding \path{SKILL.md} update is needed.} 78\% of script--Markdown co-changes in our sample are functionally coupled (RQ3).
Leaving \path{SKILL.md} behind after a script change causes Claude to invoke the skill with outdated instructions, passing flags that no longer exist, use renamed identifiers, or follow an invocation pattern that the script no longer supports, with no runtime error to alert the developer.
Developers should check their diff against the four coupling categories before submitting a script PR and include the Markdown update in the same PR rather than a follow-up.
Whether a tool can detect these four categories from a script diff and automatically propose the corresponding Markdown update remains an open problem, though the predictable structure of each category suggests it is feasible.

\section{Threats to Validity}
\label{sec:threats}
\textbf{Search completeness.}
Our discovery relies on the GitHub Code Search API, which indexes only publicly visible files.
Repositories that are private or that were deleted before our search date are not captured.
The recursive prefix partitioning mitigates the API cap but cannot recover content that is not indexed.

\textbf{Star threshold.}
The choice of ten stars as the noise-removal threshold is a tradeoff between coverage and noise removal.
We ran the full data collection pipeline at two alternative thresholds. At 5 stars: 2,694 repositories. At 10 stars (our baseline): 1,926 repositories. At 25 stars: 1,247 repositories. Across all three thresholds, Skills remains the dominant component type, \textit{feat} remains the dominant commit type, and the relative ordering of component types is consistent, supporting the robustness of our qualitative findings.

\textbf{Temporal scope.}
The dataset was collected in April 2026 and represents a snapshot of a rapidly evolving ecosystem.
Trends observed here, particularly growth rates and adoption patterns, may not generalize to later periods.
The median repository age at collection time is only 80 days, meaning many repositories had not yet accumulated a long commit history at the point of collection.

\textbf{Generalizability.}
All findings describe the Claude Code agent plugin marketplaces specifically.
Whether the observed patterns, including the SE-heavy composition, the Agents--Commands coupling, and the redefined commit type semantics, generalize to other AI coding agent platforms (e.g., Cursor, Copilot, Gemini) or to AI-native ecosystems more broadly remains untested.
Researchers should treat these findings as a baseline for Claude Code agent plugin marketplaces and verify whether they replicate before drawing conclusions about AI-native software in general.

\textbf{Manual labeling.}
Two of our analyses rely on manual coding: RQ2 classifies a stratified sample of 700 commits into content sub-categories, and RQ3 classifies a stratified sample of 64 co-change pull requests into coupling categories.
Both follow the same open coding procedure. The first two authors jointly inspect a small pilot set and construct an initial set of categories from the content and the agent plugin context, then independently code the full sample, adding a category through discussion whenever a commit or pull request does not fit an existing one, and resolving disagreements until they agree on every category definition.
Inter-rater reliability is substantial in both analyses (Cohen's $\kappa = 0.671$ across the 700 commits and $\kappa = 0.74$ across the 64 pull requests with thresholds according to \citet{landis1977measurement}), and the 13 pull-request disagreements in RQ3 were reconciled to a single agreed label.
Full validation details are reported in the RQ2 (Section~\ref{sec:rq2}) and RQ3 (Section~\ref{sec:rq3}) approach sections.
Despite these steps, ambiguous cases still require subjective judgment, for example, distinguishing an interface change from an internal logic change when a pull request does both, or an instruction-content edit from a behavioral change in a commit, and the assigned label reflects the raters' interpretation of developer intent from the diff and commit message.

\textbf{LLM-based classification.}
Plugin taxonomy classification uses Qwen3-Coder-Next-80B and commit type classification uses GPT-5-mini, and both introduce potential biases from model training data and prompt sensitivity.
We mitigate taxonomy classification risk by adopting the same scheme and model family used by Ling et al.~\cite{ling2026agentskills} for a comparable task, and commit classification risk by validating Stage~2 against human labels on two independent 385-commit samples, achieving substantial agreement ($\kappa = 0.707$).
Nonetheless, LLM outputs are non-deterministic and may differ across runs or model versions, and the classifications reported here reflect a single run on the final model versions.

\textbf{Agent detection.}
Agent contribution detection follows the heuristics of Robbes et al.~\cite{robbes2026promises}, which rely on co-author trailers, email addresses, author names, and branch prefixes.
More commonly, developers who use AI agents to generate content may commit the result themselves without adding a co-author trailer, leaving no signal for our heuristics to detect.
The 34.9\% Claude attribution rate should therefore be interpreted as a lower bound on true AI involvement rather than an exact count.

\section{Conclusion}
\label{sec:conclusion}
This paper presents the first large-scale empirical study of an AI coding agent plugin marketplace, analyzing 1,926 repositories, 8,351 plugins, and 77,773 commits from the Claude Code plugin marketplace on GitHub.
We conducted three studies to characterize the marketplace structure, its development patterns, and its component co-evolution behavior.

Our findings show that AI-native plugins are a fundamentally different class of software artifact from conventional OSS packages.
The marketplaces are young and their development is accelerating, with 84.5\% of repositories created in the six months after the official launch and plugin-touching commit activity growing 8.8$\times$ over the same period. Software Engineering plugins dominate (61.3\%), and 34.4\% of plugins combine multiple component types, providing the first empirical baseline for this class of artifact.
Development is predominantly feature-driven (\textit{feat} at 39.6\%) and heavily AI-assisted, with Claude alone co-authoring 34.9\% of all commits.
Four CCS types, \textit{docs}, \textit{perf}, \textit{style}, and \textit{refactor}, each redefine their conventional meaning in plugin repositories, where the primary artifact is a natural-language instruction file rather than executable code.
Cross-ecosystem comparisons using these labels without semantic validation risk drawing incorrect conclusions.
Only the Agents--Commands component pair co-changes above chance (Lift $= 1.40$); all other component pairs evolve independently, and Script--Markdown coupling within \path{skills/} represents a new class of maintenance dependency with no analog in traditional SE.

\bibliographystyle{ACM-Reference-Format}
\bibliography{agent_plugin}

@article{ling2026agentskills,
  title     = {Agent Skills: A Data-Driven Analysis of {Claude} Skills
               for Extending Large Language Model Functionality},
  author    = {Ling, George and Zhong, Shanshan and Huang, Richard},
  journal   = {arXiv preprint arXiv:2602.08004},
  year      = {2026}
}

@misc{claudecode_hooks,
  title        = {Automate Actions with Hooks},
  author       = {{Anthropic}},
  howpublished = {\url{https://code.claude.com/docs/en/hooks-guide}},
  year         = {2025}
}

@misc{mcp2024,
    title        = {Model Context Protocol},
    author       = {{Anthropic}},
    howpublished = {\url{https://modelcontextprotocol.io}},
    year         = {2024}
}

@misc{lsp2016,
    title        = {Language Server Protocol Specification},
    author       = {{Microsoft}},
    howpublished = {\url{https://microsoft.github.io/language-server-protocol/}},
    year         = {2016}
}

@inproceedings{li2026aidev,
  title     = {{AIDev}: Studying {AI} Coding Agents on {GitHub}},
  author    = {Li, Hao and Zhang, Haoxiang and Hassan, Ahmed E.},
  booktitle = {Proceedings of the International Conference on Mining
               Software Repositories (MSR)},
  pages     = {1029--1033},
  publisher = {{ACM}},
  year      = {2026}
}

@article{li2025rise,
  title     = {The Rise of {AI} Teammates in Software Engineering
               ({SE}) 3.0: How Autonomous Coding Agents Are Reshaping
               Software Engineering},
  author    = {Li, Hao and Zhang, Haoxiang and Hassan, Ahmed E.},
  url       = {https://doi.org/10.48550/arXiv.2507.15003},
  doi       = {10.48550/ARXIV.2507.15003},
  year      = {2025}
}

@article{li2025promptmgmt,
  title     = {Understanding Prompt Management in {GitHub}
               Repositories: {A} Call for Best Practices},
  author    = {Hao Li and
               Hicham Masri and
               Filipe Roseiro C{\^{o}}go and
               Abdul Ali Bangash and
               Bram Adams and
               Ahmed E. Hassan},
  journal   = {{IEEE} Software},
  volume    = {43},
  number    = {2},
  pages     = {85--93},
  year      = {2026},
  doi       = {10.1109/MS.2025.3644251}
}

@inproceedings{barrak2021coevolution,
  title     = {On the Co-evolution of {ML} Pipelines and Source
               Code---{Empirical} Study of {DVC} Projects},
  author    = {Barrak, Amine and Eghan, Ellis E. and Adams, Bram},
  booktitle = {Proceedings of the IEEE International Conference on
               Software Analysis, Evolution and Reengineering (SANER)},
  pages     = {422--433},
  publisher = {{IEEE}},
  year      = {2021}
}

@inproceedings{jiang2015coevolution,
  title     = {Co-evolution of Infrastructure and Source Code---{An}
               Empirical Study},
  author    = {Jiang, Yujuan and Adams, Bram},
  booktitle = {Proceedings of the International Conference on Mining
               Software Repositories (MSR)},
  pages     = {45--55},
  publisher = {{IEEE} Computer Society},
  year      = {2015}
}

@inproceedings{zeng2025ccs,
  title     = {A First Look at Conventional Commits Classification},
  author    = {Zeng, Qunhong and Zhang, Yuxia and Qiu, Zhiqing and Liu, Hui},
  booktitle = {Proceedings of the IEEE/ACM 47th International Conference on Software Engineering (ICSE)},
  pages     = {2277--2289},
  publisher = {{IEEE}},
  year      = {2025},
  doi       = {10.1109/ICSE55347.2025.00011}
}

@misc{conventionalcommits,
  title        = {Conventional Commits Specification},
  author       = {{Conventional Commits}},
  howpublished = {\url{https://www.conventionalcommits.org}},
  year         = {2024}
}

@article{kalliamvakou2016github,
  title     = {An In-Depth Study of the Promises and Perils of Mining
               {GitHub}},
  author    = {Kalliamvakou, Eirini and Gousios, Georgios and
               Blincoe, Kelly and Singer, Leif and German, Daniel M.
               and Damian, Daniela},
  journal   = {Empirical Software Engineering},
  volume    = {21},
  number    = {5},
  pages     = {2035--2071},
  year      = {2016}
}

@inproceedings{nagappan2013msr,
  title     = {Diversity in Software Engineering Research},
  author    = {Nagappan, Meiyappan and Zimmermann, Thomas and
               Bird, Christian},
  booktitle = {Proceedings of the 2013 9th Joint Meeting on Foundations of Software Engineering (ESEC/FSE 2013)},
  pages     = {466--476},
  year      = {2013},
  publisher = {ACM},
  doi       = {10.1145/2491411.2491415}
}

@article{emse22sampling,
  title        = {Sampling in Software Engineering Research:
                  {A} Critical Review and Guidelines},
  author       = {Baltes, Sebastian and Ralph, Paul},
  howpublished = {Empirical Software Engineering},
  volume       = {27},
  number       = {4},
  pages        = {94},
  year         = {2022},
  doi          = {10.1007/S10664-021-10072-8}
}

@misc{claudecode_plugins,
  title        = {Claude Code Plugins},
  author       = {{Anthropic}},
  howpublished = {\url{https://code.claude.com/docs/en/plugins}},
  year         = {2025}
}

@misc{claudecode_marketplaces,
  title        = {Claude Code Plugin Marketplaces},
  author       = {{Anthropic}},
  howpublished = {\url{https://code.claude.com/docs/en/plugin-marketplaces}},
  year         = {2025}
}

@misc{claudecode_discover,
  title        = {Discover {Claude} Code Plugins},
  author       = {{Anthropic}},
  howpublished = {\url{https://code.claude.com/docs/en/discover-plugins}},
  year         = {2025}
}

@misc{claudecode_skills,
  title        = {Extend {Claude} with Skills},
  author       = {{Anthropic}},
  howpublished = {\url{https://code.claude.com/docs/en/skills}},
  year         = {2025}
}

@inproceedings{robbes2026promises,
  author       = {Romain Robbes and
                  Th{\'{e}}o Matricon and
                  Thomas Degueule and
                  Andre Hora and
                  Stefano Zacchiroli},
  title        = {Promises, Perils, and (Timely) Heuristics for Mining Coding Agent
                  Activity},
  booktitle    = {Proceedings of the 23rd International Conference on Mining Software
                  Repositories (MSR)},
  pages        = {496--507},
  publisher    = {{ACM}},
  year         = {2026},
  doi          = {10.1145/3793302.3793375}
}

@misc{wan2025commitsuite,
  title     = {{CommitSuite}: A Comprehensive Benchmark for Commit Classification and Message Generation},
  author    = {Wan, Zirui and Wu, Zhaonan and Hou, Xinyi and Zhao, Yanjie and Xia, Pengcheng and Wang, Haoyu},
  year      = {2025},
  note      = {\url{https://arxiv.org/abs/2605.02256}}
}

@inproceedings{swanson1976dimensions,
  title     = {The dimensions of maintenance},
  author    = {Swanson, E. Burton},
  booktitle = {Proceedings of the 2nd International Conference on Software Engineering},
  pages     = {492--497},
  year      = {1976}
}

@article{mann1947test,
  title     = {On a test of whether one of two random variables is stochastically larger than the other},
  author    = {Mann, Henry B. and Whitney, Donald R.},
  journal   = {The Annals of Mathematical Statistics},
  volume    = {18},
  number    = {1},
  pages     = {50--60},
  year      = {1947}
}

@article{decan2018empirical,
  author  = {Alexandre Decan and Tom Mens and Philippe Grosjean},
  title   = {An Empirical Comparison of Dependency Network Evolution
             in Seven Software Packaging Ecosystems},
  journal = {Empirical Software Engineering},
  volume  = {24},
  number  = {1},
  pages   = {381--416},
  year    = {2019},
  doi     = {10.1007/S10664-017-9589-Y}
}

@inproceedings{kikas2017structure,
  author    = {Riivo Kikas and Georgios Gousios and Marlon Dumas
               and Dietmar Pfahl},
  title     = {Structure and Evolution of Package Dependency Networks},
  booktitle = {Proceedings of the 14th International Conference on
               Mining Software Repositories (MSR)},
  pages     = {102--112},
  publisher = {{IEEE} Computer Society},
  year      = {2017},
  doi       = {10.1109/MSR.2017.55}
}

@article{watanabe2025agentic,
  author  = {Miku Watanabe and Hao Li and Yutaro Kashiwa and
             Brittany Reid and Hajimu Iida and Ahmed E. Hassan},
  title   = {On the Use of Agentic Coding: An Empirical Study of
             Pull Requests on {GitHub}},
  journal = {ACM Transactions on Software Engineering and Methodology (TOSEM)},
  publisher = {{ACM}},
  year    = {2026},
  doi     = {10.1145/3798166},
}

@article{ouatiti2026logging,
  author  = {Youssef Esseddiq Ouatiti and Mohammed Sayagh and
             Hao Li and Ahmed E. Hassan},
  title   = {Do {AI} Coding Agents Log Like Humans? An Empirical
             Study},
  journal = {arXiv preprint arXiv:2604.09409},
  year    = {2026},
  note    = {\url{https://arxiv.org/abs/2604.09409}}
}

@inproceedings{tufano2024chatgpt,
  author    = {Rosalia Tufano and Antonio Mastropaolo and Federica Pepe
               and Ozren Dabi\'{c} and Massimiliano {Di Penta}
               and Gabriele Bavota},
  title     = {Unveiling {ChatGPT's} Usage in Open Source Projects:
               {A} Mining-Based Study},
  booktitle = {Proceedings of the 21st International Conference on
               Mining Software Repositories (MSR)},
  year      = {2024},
  pages     = {571--583},
  publisher = {{ACM}},
  doi       = {10.1145/3643991.3644918}
}

@inproceedings{zaidman2008coevolution,
  author    = {Andy Zaidman and Bart {Van Rompaey} and Serge Demeyer
               and Arie van Deursen},
  title     = {Mining Software Repositories to Study Co-Evolution of
               Production \& Test Code},
  booktitle = {Proceedings of the 2008 International Conference on
               Software Testing, Verification, and Validation (ICST)},
  pages     = {220--229},
  publisher = {{IEEE} Computer Society},
  year      = {2008},
  doi       = {10.1109/ICST.2008.47}
}

@inproceedings{fluri2007comments,
  author    = {Beat Fluri and Michael W\"{u}rsch and Harald C. Gall},
  title     = {Do Code and Comments Co-Evolve? {On} the Relation
               Between Source Code and Comment Changes},
  booktitle = {Proceedings of the 14th Working Conference on Reverse Engineering (WCRE)},
  year      = {2007},
  pages     = {70--79},
  publisher = {{IEEE} Computer Society},
  doi       = {10.1109/MSR.2007.5}
}

@article{horikawa2025refactoring,
  author  = {Kosei Horikawa and Hao Li and Yutaro Kashiwa and Bram Adams
             and Hajimu Iida and Ahmed E. Hassan},
  title   = {Agentic Refactoring: An Empirical Study of {AI} Coding Agents},
  journal = {arXiv preprint arXiv:2511.04824},
  year    = {2025},
  note    = {\url{https://arxiv.org/abs/2511.04824}}
}

@article{honel2020density,
  author    = {Sebastian H\"{o}nel and Morgan Ericsson and Welf L\"{o}we
             and Anna Wingkvist},
  title     = {Using Source Code Density to Improve the Accuracy of
             Automatic Commit Classification into Maintenance Activities},
  journal   = {Journal of Systems and Software},
  volume    = {168},
  pages     = {110673},
  year      = {2020},
  publisher = {Elsevier}
}

@inproceedings{yan2024chatgpt,
  author    = {Chuan Yan and Ruomai Ren and Mark Huasong Meng and
               Liuhuo Wan and Tian Yang Ooi and Guangdong Bai},
  title     = {Exploring {ChatGPT} App Ecosystem: Distribution, Deployment and Security},
  booktitle = {Proceedings of the 39th IEEE/ACM International Conference
               on Automated Software Engineering (ASE)},
  pages     = {1370--1382},
  publisher = {{ACM}},
  year      = {2024},
  doi       = {10.1145/3691620.3695510}
}

@inproceedings{su2025gpt,
  author    = {Dongxun Su and Yanjie Zhao and Xinyi Hou and
               Shenao Wang and Haoyu Wang},
  title     = {{GPT} Store Mining and Analysis},
  booktitle = {Proceedings of the 16th International Conference on Internetware},
  pages     = {344--354},
  publisher = {{ACM}},
  year      = {2025},
  doi       = {10.1145/3755881.3755900}
}

@article{onagh2025extension,
  author    = {Elnaz Onagh and Meiyappan Nayebi},
  title     = {Extension Decisions in Open Source Software Ecosystem},
  journal   = {Journal of Systems and Software},
  volume    = {230},
  pages     = {112552},
  year      = {2025},
  doi       = {10.1016/J.JSS.2025.112552}
}

@inproceedings{hassan2024fmware,
  author       = {Ahmed E. Hassan and
                  Dayi Lin and
                  Gopi Krishnan Rajbahadur and
                  Keheliya Gallaba and
                  Filipe Roseiro C{\^{o}}go and
                  Boyuan Chen and
                  Haoxiang Zhang and
                  Kishanthan Thangarajah and
                  Gustavo Ansaldi Oliva and
                  Jiahuei (Justina) Lin and
                  Wali Mohammad Abdullah and
                  Zhen Ming (Jack) Jiang},
  title     = {Rethinking Software Engineering in the Era of Foundation Models:
               {A} Curated Catalogue of Challenges in the Development of
               Trustworthy {FMware}},
  booktitle = {Companion Proceedings of the 32nd {ACM} International Conference
               on the Foundations of Software Engineering ({FSE})},
  pages     = {294--305},
  publisher = {{ACM}},
  year      = {2024},
  doi       = {10.1145/3663529.3663849}
}

@inproceedings{marsavina2014coevolution,
  author    = {Cosmin Marsavina and Daniele Romano and Andy Zaidman},
  title     = {Studying Fine-Grained Co-Evolution Patterns of Production
               and Test Code},
  booktitle = {Proceedings of the 14th {IEEE} International Working Conference
               on Source Code Analysis and Manipulation ({SCAM})},
  pages     = {195--204},
  publisher = {{IEEE} Computer Society},
  year      = {2014},
  doi       = {10.1109/SCAM.2014.6975653}
}

@inproceedings{ait2022survival,
  author    = {Ait, Adem and C{\'a}novas Izquierdo, Javier Luis and Cabot, Jordi},
  title     = {An empirical study on the survival rate of {GitHub} projects},
  booktitle = {Proceedings of the 19th International Conference on Mining Software Repositories (MSR)},
  year      = {2022},
  pages     = {365--375},
  publisher = {{ACM}},
  doi       = {10.1145/3524842.3527941}
}

@article{wen2024multilingual,
  author    = {Wen Li and Austin Marino and Haoran Yang and Na Meng and Li Li and Haipeng Cai},
  title     = {How Are Multilingual Systems Constructed: Characterizing Language Use and Selection in Open-Source Multilingual Software},
  journal   = {{ACM} Transactions on Software Engineering and Methodology},
  volume    = {33},
  number    = {3},
  pages     = {63:1--63:46},
  year      = {2024},
  doi       = {10.1145/3631967}
}

@article{li2025bridging,
  author    = {Hao Li and Cor-Paul Bezemer},
  title     = {Bridging the language gap: an empirical study of bindings for open source machine learning libraries across software package ecosystems},
  journal   = {Empirical Software Engineering},
  volume    = {30},
  number    = {1},
  pages     = {6},
  year      = {2025},
  publisher = {Springer},
  doi       = {10.1007/s10664-024-10570-5}
}

@inproceedings{zhu2026skillclone,
  title     = {{SkillClone}: Multi-Modal Clone Detection and Clone
               Propagation Analysis in the Agent Skill Ecosystem},
  author    = {Zhu, Jiaying and Zhang, Lyuye and Guo, Wenbo and Liu, Yang},
  journal      = {CoRR},
  volume       = {abs/2603.22447},
  year         = {2026},
  url          = {https://doi.org/10.48550/arXiv.2603.22447},
  doi          = {10.48550/ARXIV.2603.22447}
}

@article{dig2006apis,
  author  = {Dig, Danny and Johnson, Ralph},
  title   = {How do {APIs} evolve? {A} story of refactoring},
  journal = {Journal of Software Maintenance and Evolution: Research and Practice},
  volume  = {18},
  number  = {2},
  pages   = {83--107},
  year    = {2006},
  doi     = {10.1002/SMR.328}
}

@book{lune2017qualitative,
  title={Qualitative research methods for the social sciences},
  author={Lune, Howard and Berg, Bruce L},
  year={2017},
  publisher={Pearson}
}

@article{cliff1993dominance,
  title={Dominance statistics: Ordinal analyses to answer ordinal questions.},
  author={Cliff, Norman},
  journal={Psychological bulletin},
  volume={114},
  number={3},
  pages={494},
  year={1993},
  publisher={American Psychological Association}
}

@article{wilcoxon1945individual,
  title={Individual comparisons by ranking methods},
  author={Wilcoxon, Frank},
  journal={Biometrics bulletin},
  volume={1},
  number={6},
  pages={80--83},
  year={1945},
  publisher={JSTOR}
}

@article{mann1945nonparametric,
  title={Nonparametric tests against trend},
  author={Mann, Henry B},
  journal={Econometrica: Journal of the econometric society},
  pages={245--259},
  year={1945},
  publisher={JSTOR}
}

@article{kendall1962rank,
  title={Rank correlation methods},
  author={Kendall, Maurice George and Gibbons, Jean Dickinson},
  year={1962},
  publisher={Griffin London}
}

@article{sen1968estimates,
  title={Estimates of the regression coefficient based on Kendall's tau},
  author={Sen, Pranab Kumar},
  journal={Journal of the American statistical association},
  volume={63},
  number={324},
  pages={1379--1389},
  year={1968},
  publisher={Taylor \& Francis}
}

@inproceedings{kudrjavets2023codevelocity,
  author    = {Kudrjavets, Gunnar and Nagappan, Nachiappan and Rastogi, Ayushi},
  title     = {Are We Speeding Up or Slowing Down? On Temporal Aspects of Code Velocity},
  booktitle = {2023 IEEE/ACM 20th International Conference on Mining Software Repositories (MSR)},
  pages     = {267--271},
  year      = {2023},
  publisher = {{IEEE}},
  doi       = {10.1109/MSR59073.2023.00046}
}

@conference{romano2006appropriate,
  title     = {Appropriate Statistics for Ordinal Level Data: Should We Really Be Using T-Test and Cohen'd for Evaluating Group Differences on the NSSE and Other Surveys},
  author    = {Romano, Jeanine and Kromrey, Jeffrey D and Coraggio, Jesse and Skowronek, Jeff},
  booktitle = {annual meeting of the Florida Association of Institutional Research},
  pages     = {1--3},
  year      = {2006}
}

@inproceedings{hindle2008large,
  author    = {Abram Hindle and Daniel M. German and Ric Holt},
  title     = {What Do Large Commits Tell Us? {A} Taxonomical Study of Large Commits},
  booktitle = {Proceedings of the 5th International Working Conference on
               Mining Software Repositories (MSR)},
  pages     = {99--108},
  year      = {2008},
  doi       = {10.1145/1370750.1370773}
}

@article{bhatia2023towards,
  author    = {Aaditya Bhatia and Ellis E. Eghan and Manel Grichi and
               William G. Cavanagh and Zhen Ming Jiang and Bram Adams},
  title     = {Towards a Change Taxonomy for Machine Learning Pipelines:
               Empirical Study of {ML} Pipelines and Forks Related to
               Academic Publications},
  journal   = {Empirical Software Engineering},
  volume    = {28},
  number    = {3},
  pages     = {60},
  year      = {2023},
  doi       = {10.1007/s10664-022-10282-8}
}

@inproceedings{mcintosh2011build,
  author    = {Shane McIntosh and Bram Adams and Thanh H. D. Nguyen and Yasutaka Kamei and Ahmed E. Hassan},
  title     = {An Empirical Study of Build Maintenance Effort},
  booktitle = {Proceedings of the 33rd International Conference on Software Engineering (ICSE)},
  pages     = {141--150},
  year      = {2011},
  doi       = {10.1145/1985793.1985813}
}

@article{landis1977measurement,
  title={The measurement of observer agreement for categorical data},
  author={Landis, J Richard and Koch, Gary G},
  journal={biometrics},
  pages={159--174},
  year={1977},
  publisher={JSTOR}
}

\appendix

\definecolor{framecolor}{RGB}{160, 0, 0}
\definecolor{framecolorbg}{RGB}{255, 250, 250}
\definecolor{titlecolorbg}{RGB}{255, 240, 240}
\newtcblisting{appendixlstbox}[1]{
    enhanced,
    colback=framecolorbg,
    colbacktitle=titlecolorbg,
    coltitle=black,
    colframe=framecolor,
    fonttitle=\bfseries,
    listing only,
    title=\textbf{#1},
    breakable,
    listing options={
        basicstyle=\ttfamily\small,
        breaklines=true,
        columns=fullflexible
    }
}

\section{Plugin Classification Prompt}
\label{app:classification_prompt}        
Listing~\ref{lst:classification_prompt} shows the prompt used to classify plugins into the 6 major categories and 20 sub-categories with Qwen3-Coder-Next-80B.

\begin{appendixlstbox}{Prompt for Plugin Use Case Classification}
Role
You are an expert AI Agent Plugin Classifier. Your task is to categorize a given agent plugin into a specific taxonomy based on its name and description.

Taxonomy
You must classify the plugin into one of the following 6 Major Categories and their corresponding Sub Categories. Read the definitions carefully.

1. Software Engineering
1.1 Code Generation: Plugins related to writing source code, generating unit tests, code translation, refactoring, or code completion.
1.2 Debug & Analysis: Plugins for finding bugs, static analysis, code explanation, security auditing, or linting.
1.3 Version Control: Plugins involving Git, GitHub, GitLab, managing pull requests, commits, or branching.
1.4 Infrastructure: Plugins related to DevOps, Python environment, cloud services (AWS/Azure), Docker, Kubernetes, CI/CD pipelines, or server deployment.
2. Information Retrieval
2.1 Web Search: General purpose internet search engines (Google/Bing) for current events, general knowledge, or news.
2.2 Academic Search: Searching specific knowledge bases, encyclopedias (Wikipedia), academic papers (ArXiv), or legal databases.
2.3 Live Data Streams: Fetching real time dynamic data such as stock prices, weather forecasts, traffic status, or sports scores.
3. Productivity Tools
3.1 Team Communication: Tools for messaging (Slack/Discord), emails, calendar scheduling, or meeting management.
3.2 Document Systems: Interactions with documentation tools (Notion/Google Docs), wiki systems, or reading/parsing PDF documents.
3.3 Task Management: Plugins for project planning tools (Jira/Trello), to do lists, or issue tracking.
4. Data & Analytics
4.1 Data Processing: ETL tasks, data cleaning, format conversion (JSON to CSV), sorting, filtering, or database querying (SQL).
4.2 Math & Calculation: Performing mathematical operations, using calculators, symbolic math, or complex physics/logic formulas.
4.3 Data Visualization: Generating charts, graphs, plots, or visual reports from data sets.
5. Content Creation
5.1 Image Generation: Creating images from text, editing photos, style transfer, or object removal.
5.2 Text Generation: Creative writing, storytelling, translation between languages, poetry, or marketing copy.
5.3 Audio & Video: Text to Speech, Speech to Text, video editing, music generation, or video analysis.
6. Utilities & Other
6.1 Local File Control: Operations on the local file system such as reading, writing, moving, or deleting files and folders.
6.2 Command Execution: Running shell commands, terminal operations, or monitoring system resources (CPU/RAM).
6.3 Memory & Cognition: Managing conversation history, summarizing long contexts for memory, or storing user preferences.
6.4 Other Utilities: Miscellaneous tools that do not fit elsewhere, such as random number generation, UUID creation, or specific API wrappers.

Constraints
Analyze the Plugin Name and Plugin Description deeply.
Select exactly ONE Sub Category that best fits the plugin.
Output the result in strict JSON format.
Do not output any conversational text.

Input Format
Plugin Name: [Name] Plugin Description: [Description]

Output Format

"major_category_id": "Number",
"major_category_name": "String",
"sub_category_id": "Number.Number",
"sub_category_name": "String",
"reasoning": "Brief explanation in English"

Task
Input: Plugin Name: PLUGIN_NAME Plugin Description: PLUGIN_DESCRIPTION

Output:
\end{appendixlstbox}
\begin{minipage}{\linewidth}
\captionof{lstlisting}{Prompt for plugin use case classification (Qwen3-Coder-Next-80B)}
\label{lst:classification_prompt}
\end{minipage}

\section{Conventional Commit Type Classification Prompt}
\label{app:ccs_prompt}
Listing~\ref{lst:ccs_prompt} shows the prompt used to classify the 26,632 non-conforming commits into the 12-category CCS taxonomy with GPT-5-mini.
Each commit is accompanied by contextual metadata: repository name, commit hash, message, plugin component types touched, file counts, and plugin file ratio.

\begin{appendixlstbox}{Prompt for Conventional Commit Classification}
You are a Conventional Commit classifier. Classify each commit into exactly one type.

Use the commit message as the primary signal. You may also use changed-file context and metadata to disambiguate whether the work is documentation, tests, CI/build, refactoring, chores, or feature/bug work.
Do not invent missing details. If the evidence is weak, choose the best-fitting label and lower confidence.

Types:
- feat: A new feature
- fix: A bug fix
- docs: Documentation only changes
- style: Changes that do not affect the meaning of the code
- refactor: A code change that neither fixes a bug nor adds a feature
- perf: A code change that improves performance
- test: Adding missing tests or correcting existing tests
- build: Changes that affect the build system or external dependencies
- ci: Changes to our CI configuration files and scripts
- chore: Changes to the build process or auxiliary tools
- revert: Reverts a previous commit
- other: Any other changes that do not fit the above categories

Return valid JSON with this schema:
{"results":[{"id": <int>, "output": "<type>",
  "reason": "<brief reason>", "confidence": <1-10>}]}

--- Per-commit context (provided for each commit) ---
COMMIT: <repo> @ <hash>
Message: <commit message>
Components: <plugin component types touched>
Files changed: <total> total (<plugin> plugin, <non-plugin> non-plugin)
Plugin ratio: <ratio>
\end{appendixlstbox}
\begin{minipage}{\linewidth}
\captionof{lstlisting}{Prompt for Conventional Commit classification (GPT-5-mini)}
\label{lst:ccs_prompt}
\end{minipage}

\section{Commit Re-classification Prompt}
\label{app:relabel_prompt}
Listing~\ref{lst:relabel_prompt} shows the prompt used to re-classify commits
originally labeled \textit{docs}, \textit{ci}, or \textit{style} into the
12-category CCS taxonomy with GPT-5-mini. A type-specific rubric is selected by
the commit's original label, and each commit is provided with its message and a
per-file sample of its diff, grouped by file role (AI-read instruction files,
human-facing documentation, configuration, workflows, and scripts).

\begin{appendixlstbox}{Prompt for commit re-classification (rubric selected by original label)}
--- rubric for `docs` ---
You are re-classifying a commit a developer labeled `docs` in an AI-agent PLUGIN repo. Decide its TRUE Conventional Commit type from the DIFF.

AI-READ files (SKILL.md, agents/*.md, commands/*.md, skills/**, references/*.md) are the agent's runtime instructions; changing their CONTENT changes behavior. HUMAN-READ files (README, CONTRIBUTING, CHANGELOG, and archived / context / userstory files) are ignored by the agent.

Judge by what the +/- lines DO:
- feat  -> ADDS an instruction the agent will act on (new section, rule, step, capability, procedure, example), OR registers a NEW plugin/component entry in marketplace.json / plugin.json.
- fix   -> CORRECTS existing agent behavior: a wrong tool/step/path/example, a rule, a script-invocation procedure, or a skill's `description` / frontmatter TRIGGER (which controls WHEN the skill fires).
- chore -> ONLY version bumps, or REMOVING / reorganizing marketplace or plugin entries and metadata (not adding new ones).
- docs  -> behavior unchanged: only human-facing or archived/context files; OR, even inside an AI-read file, only inert metadata (version, date, badge) or a LINK change (adding, removing, or fixing a source-attribution or bare reference link).

Read the DIFF, not the message. Key distinctions learned from real cases:
- Removing/fixing a source-attribution or dead link, or adding a bare reference/doc link, is informational -> docs (even in a SKILL.md).
- Adding a new plugin entry to marketplace.json -> feat; removing one or editing its metadata -> chore.
- Changing a skill's `description`/trigger, or the procedure the agent uses to invoke a script -> fix.
- A SKILL.md touched only by a version bump or whitespace -> docs; the real change may sit in a README while the AI-read file only got a bump -> docs.

--- rubric for `ci` ---
You are re-classifying a commit a developer labeled `ci` in an AI-agent PLUGIN repo. `ci:` is sometimes a real pipeline change, sometimes just a TOPIC prefix for a CI-themed plugin. Decide from the DIFF.

- ci   -> the diff changes an ACTUAL pipeline/CI automation: .github/workflows/*, GitLab/CircleCI configs, release or CI-validation scripts.
- feat -> the diff builds or extends a CI-EXPERT agent/skill/command (AI-read instruction files); no pipeline file changed.
- fix  -> the diff corrects existing CI-expert instruction content; no pipeline file changed.

Read the DIFF, not the message. No workflow/CI-config file touched and the change is to SKILL.md/agents/commands -> feat/fix, not ci (the plugin's TOPIC is CI, but nothing ran). A real workflow file changed -> ci, even if plugin files were also touched.

--- rubric for `style` ---
You are re-classifying a commit a developer labeled `style` in an AI-agent PLUGIN repo. Traditional style = formatting with no change in meaning; but rewording an AI-read instruction can change agent behavior, which is not style. Decide from the DIFF.

- style -> pure formatting, no change in meaning: whitespace, backticks, markdown structure, script linting/reformatting, or reordering that changes no instruction.
- feat  -> the wording change adds or strengthens a directive so the agent does something new/different, or changes the verbatim text the agent outputs.
- fix   -> the wording change corrects an existing directive so the agent behaves correctly.

Read the DIFF, not the message. Ask: would the agent behave differently after this change? Yes -> feat/fix. Cosmetic only -> style. Reformatting .py/.sh/.js is style.

--- output format (appended to the selected rubric) ---
true_type may be ANY of: feat, fix, docs, style, refactor, perf, test, build, ci, chore, revert, other. Keeping <original label> is fine if it still fits. Most re-labels are feat/fix; let the diff decide.
Return STRICT JSON only:
{"true_type":"<type>","moved":<true|false>,"reason":"<short: what changed and why>","confidence":<1-10>}

--- Per-commit input ---
Original label: <docs|ci|style>
Repo: <repo> @ <hash>
Files changed by role: <role:count ...>
Commit message (may be unreliable): <message, truncated>
FULL DIFF (grouped by file role): <per-file sample of the diff>
\end{appendixlstbox}
\begin{minipage}{\linewidth}
\captionof{lstlisting}{Prompt for commit re-classification (GPT-5-mini)}
\label{lst:relabel_prompt}
\end{minipage}

\section{Script--Markdown Coupling Classification Prompt}
\label{app:coupling_prompt}
Listing~\ref{lst:coupling_prompt} shows the full prompt used to classify the 259
co-change pull requests into the coupling taxonomy with \texttt{gpt-5-mini}. Each pull
request is presented to the model as its full script and Markdown diff. The validation
code is available in the replication package.

\begin{appendixlstbox}{Prompt for Script--Markdown Coupling Classification}
You are labeling a merged pull request from an AI-agent PLUGIN repository.

Every PR you see modifies BOTH:
  (1) an implementation SCRIPT (.py, .sh, or .ts), and
  (2) a natural-language INSTRUCTION file (SKILL.md, or a supporting .md) in the SAME
      skills/ subdirectory.
The instruction file is what the AI agent READS AT RUNTIME to learn how to invoke and
use the script, so a change to the script often forces the instruction file to follow.

Label the PR in TWO steps. Read the DIFFS, not any message.

========================================================================
STEP 1 -- IS IT COUPLED? (do this first; most mistakes happen here)
========================================================================
Coupling is a relationship between the SCRIPT change and the MARKDOWN change: the
markdown edit must CLOSELY REFLECT the SPECIFIC thing the script changed. Both files
changing in the same PR is NOT enough on its own.

Ask: does the markdown edit closely reflect THIS script edit?
  - YES -- the markdown documents or updates the EXACT flag, value, path, behavior, or
    output that the script changed -> it is COUPLED, go to STEP 2.
  - NO -> answer no_coupling.

Choose no_coupling whenever ANY of these hold:
  - The script changed, but the markdown does NOT closely reflect that specific change
    (the markdown edit is about something else, or never mentions what the script did).
  - Either side is cosmetic only: whitespace, emoji->ASCII, table alignment, backticks,
    reordering, reformatting with no change in meaning.
  - The markdown edit addresses a DIFFERENT concern than the script edit (e.g. the
    script fixes a regex while the markdown standardizes an invocation path).
  - The markdown documents behavior that ALREADY existed, or is a general cleanup,
    trimming, or reorganization not driven by the script edit.
  - The script and markdown edits are in DIFFERENT sub-skills bundled in one PR.
A script's BEHAVIOR changing does NOT by itself make the doc coupled -- the doc must
closely reflect THAT change. Be willing to answer no_coupling.

no_coupling examples:
  - Script replaces emoji with ASCII in echo lines; SKILL.md replaces the same emoji in
    bullets. Cosmetic on both sides. => no_coupling
  - Script adds a --limit option; the markdown adds unrelated sections and never
    documents --limit. The specific change is not reflected. => no_coupling
  - Script tightens a regex / error handling; the markdown only adds a "run from repo
    root, use ${CLAUDE_SKILL_ROOT}" note. Different concern. => no_coupling
  - A test is added for behavior that already existed; the doc note describes that
    pre-existing behavior. => no_coupling

========================================================================
STEP 2 -- IF COUPLED, WHICH MECHANISM? (pick exactly one)
========================================================================
First split into SURFACE vs CAPABILITY, then pick the leaf.

Q: did the skill's CAPABILITIES or BEHAVIOR change, or did only a NAME/VALUE/FILE that
   the markdown quotes change (skill does the same thing)?

--- SURFACE change (same capabilities; only a quoted string moved) ---
value_version_sync -> the moved string is a VALUE the skill PASSES or EMBEDS: a version,
  a default value, a MODEL identifier, an ENVIRONMENT-VARIABLE NAME, a token/field id, or
  the skill's OWN `name` identity.
repo_restructuring -> the moved string is a FILE name, DIRECTORY, PATH, an invoked
  BINARY name, or a CONFIG-FILE FORMAT (json->yaml, json->.env, moved storage/hook path,
  renamed data/reference files).
  (A renamed FILE/PATH/BINARY/FORMAT is repo, NOT value. The skill's own `name` identity
   is value, NOT repo.)

--- CAPABILITY / BEHAVIOR change --- decide by the OBSERVABLE MARKDOWN SIGNAL:
interface_change -> the markdown ADDED or REMOVED a CALLABLE ENTRY: a new `--flag`/option
  row in a parameters table, a new subcommand/command in a usage or commands list, or it
  DELETED one. The list of things the agent can TYPE grew or shrank. (Also: the calling
  convention changed, e.g. positional args -> one JSON blob.)
internal_logic_change -> the markdown did NOT add/remove a callable entry; it changed
  PROSE, EXAMPLES, or OUTPUT descriptions: documenting auto-detection (an existing arg
  becomes optional), a NEW OUTPUT field, a changed output schema, RENAMED OUTPUT field
  names, added validation, changed semantics, or a stub becoming a real implementation.

PRECEDENCE when a PR changes several things at once (this resolves most hard cases):
An INPUT-SURFACE change WINS over an accompanying behavior/output change. In order:
  1. a NEW input flag / command / subcommand ADDED or REMOVED -> interface_change
     (even if behavior also changed)
  2. a renamed INPUT value the agent passes (model id, env-var NAME, version, default,
     the skill's own `name`) -> value_version_sync (even if behavior also changed)
  3. a renamed FILE / PATH / invoked BINARY / config-file FORMAT -> repo_restructuring
     (even if behavior also changed)
  4. ONLY if there is NO input-surface change above -- the change is purely OUTPUT (a
     new or renamed output field, a changed schema), or AUTO-DETECTION (an existing arg
     becomes optional), or added validation, or a stub becoming real -> internal_logic_change
NOTE: an argument becoming OPTIONAL via auto-detection is NOT an input-surface change; it
is internal_logic_change. A new OUTPUT field is NOT an input-surface change either.

CRITICAL edge cases (these are exactly where labelers disagreed -- follow them):
- RENAMED OUTPUT field names that change the result contract (hits/misses -> assertions;
  time_to_close -> time_to_resolve) are internal_logic_change (the output contract
  changed), NOT value_version_sync. value_version_sync is only for INPUT values the agent
  passes (model id, env-var name, version, default).
- A NEW subcommand shown as a new command block (e.g. a `console` command) is
  interface_change, even if it also collects or returns data.
- An argument becoming OPTIONAL because the script now auto-detects it is
  internal_logic_change (behavior), NOT interface_change.
- Renaming the skill's OWN `name` identity -> value_version_sync; renaming DATA/REFERENCE
  files -> repo_restructuring.

WORKED EXAMPLES (coupled)
  interface_change:
  - The markdown adds two new rows --image-field and --skip-fields to its parameter table
  - The markdown adds a new `console` command block to its commands list (script added a
    new subcommand), even though the command also returns collected data
  - The markdown removes an API-endpoint / method row that the script deleted
  - The markdown rewrites every usage line from positional args to one JSON argument
  internal_logic_change:
  - The markdown documents that a command now auto-detects its dir, so an argument is
    optional; no flag/command row was added or removed
  - The markdown adds a prose section describing a NEW OUTPUT field (a `trends` object)
  - The markdown updates the output contract, renaming result fields hits/misses ->
    assertions and the score formula
  - The markdown documents new validation of a required response field
  - The markdown is rewritten from a stub description to real usage (script stub -> real)
  value_version_sync:
  - A model default gpt-5.2-mini -> claude-haiku-4-5, quoted across many docs, updated
  - An env-var NAME JIRA_AUTH_TOKEN -> JIRA_API_TOKEN, or opaque field ids swapped
  - A version string bumped; the skill's own `name:` renamed
  repo_restructuring:
  - A config file format .alva.json (JSON) -> .env (KEY=VALUE); docs replace the examples
  - Data/reference FILES renamed (ecosystem-research.md -> spring-boot-ecosystem-research.md)
  - The invoked CLI binary renamed agent-channel -> agent-channeltalk; a storage PATH moved

Return STRICT JSON only, no prose. Give a RANKED list of labels, best first:
{"labels":["<primary label>", "<optional second label>"],
 "reason":"<short: coupled? which mechanism(s), and if two, why this order>",
 "confidence":<1-10>}

HOW MANY LABELS:
- Give ONE label when a single mechanism clearly drives the Markdown edit. This is the
  common case.
- Give TWO ranked labels ONLY when the pull request genuinely does two things at once and
  BOTH drive the Markdown edit, so that dropping either one would misrepresent the change
  (for example: a NEW FLAG is added AND a model/default VALUE is changed, and the Markdown
  documents both). Rank the stronger driver first.
- Do NOT add a second label to hedge, to play safe, or when one mechanism clearly
  dominates. If one label is the honest answer, give one. Never list more than two.
Every label must be one of: interface_change, internal_logic_change, value_version_sync,
repo_restructuring, no_coupling. The FIRST label is your single best answer.
\end{appendixlstbox}
\begin{minipage}{\linewidth}
\captionof{lstlisting}{Prompt for Script--Markdown coupling classification (\texttt{gpt-5-mini}).}
\label{lst:coupling_prompt}
\end{minipage}

\end{document}